\documentclass[a4paper, 10pt, aps, twocolumn]{revtex4}
\usepackage{graphicx}
\usepackage{amsfonts}
\usepackage{amssymb}
\usepackage{amsbsy}
\usepackage{amsmath}
\usepackage{mathrsfs}
\usepackage{latexsym}
\usepackage{bm}
\usepackage{subfigure} 
\usepackage{wasysym}
\usepackage{mathbbol}
\usepackage{bigints} 
\usepackage{fontawesome}
\usepackage{natbib}
\usepackage{color}

\usepackage{footnote}
\usepackage[colorlinks = true, linkcolor=blue, citecolor=blue, urlcolor=cyan]{hyperref}

\def\cal{\mathcal{}}
\def\red{\color{red}}

\begin{document}

\title[Accretion in deformed Kerr spacetime]{Properties of accretion flow in deformed Kerr spacetime}

\author{Subhankar Patra}
\email{psubhankar@iitg.ac.in}

\author{Bibhas Ranjan Majhi}
\email{bibhas.majhi@iitg.ac.in}

\author{Santabrata Das}
\email{sbdas@iitg.ac.in}

\affiliation{Department of Physics, Indian Institute of Technology Guwahati, 
	Guwahati 781039, Assam, India}

\date{\today}

\begin{abstract}
	We study the properties of a low-angular momentum, inviscid, advective accretion flow in a deformed Kerr spacetime under the framework of general theory of relativity. We solve the governing equations that describe the flow motion in terms of input parameters, namely energy ($E$), angular momentum ($\lambda$), spin ($a_{\rm k}$) and deformation parameter ($\varepsilon$), respectively. We find that global transonic accretion solutions continue to exist in non-Kerr spacetime. Depending on the input parameters, accretion flow is seen to experience shock transition and we find that shocked induced accretion solutions are available for a wide range of the parameter space in $\lambda-E$ plane. We examine the modification of the shock parameter space with $\varepsilon$, and find that as $\varepsilon$ is increased, the effective region of the parameter space is reduced, and gradually shifted towards the higher $\lambda$ and lower $E$ domain. In addition, for the first time in the literature, we notice that accretion flow having zero angular momentum admits shock transition when spacetime deformation is significantly large. Interestingly, beyond a critical limit of $\varepsilon^{\rm max}$, the nature of the central object alters from black hole (BH) to naked singularity (NS) and we identify $\varepsilon^{\rm max}$ as function of $a_{\rm k}$. Further, we examine the accretion solutions and its properties around the naked singularity as well. Finally, we indicate the implications of the present formalism in the context of astrophysical applications.
\end{abstract}




\maketitle
  
\section{Introduction}

Accretion of matter onto compact objects (namely, black holes (BHs) and neutron stars) is the most acceptable and prolific physical process in the context of energy generation in enigmatic objects like active galactic nuclei (AGNs) and X-ray binaries (XRBs) etc \cite[]{Pringle_1981, Frank_2002, netzer_2013, Abramowicz_2013}. It has been noticed from observations that these objects undergo several spectral state transitions \cite[]{esin_et_al_1998}, and these spectral states are classified as ``Low/Hard'' state (LHS), ``High/Soft'' state (HSS), and ``Intermediate'' state (IMS), respectively. In order to understand the aforementioned spectral states, various accretion disc models have been developed. The standard thin disc model, developed considering Keplerian flow, \cite[]{Shakura_Sunyaev_1973, Novikov_Thorne_1973} was successful to explain the HSS. To interpret the characteristics of LHS and IMS, several other disc models, namely the thick disc model \cite[]{paczynsky1980thick, Paczy_Wiita_1980, Chakrabarti_1990_thick_disc_issue, Molteni_Lanzafame_Chakrabarti_1994}, advective disc model \cite[]{Liang_Thompson_1980, Fukue_1987,Chakrabarti_1989}, advection-dominated disc model \cite[]{Narayan_Yi_1994, Chakrabarti_1996,Esin_et_al_1997, narayan_et_al_1997, Lu_Gu_Yuan_1999} and truncated disc models \cite[]{Esin_et_al_2001, done_2007} were studied. Among them, some of the disc models are also potentially viable to comprehend the origin of the relativistic jets and quasi-periodic oscillation (QPOs) phenomena as well. Needless to mention that the underlying scenario of these theories are landed into the fundamental aspects of the general relativity (GR). Since the hosted central object directly impacts on the properties of the accretion disc, such theory can successfully probe the signatures of strong gravity ($e.g.$, event horizon, ergosphere, ISCO and shadow etc.), and eventually, one can ascertain the physical parameters ($i.e.$, mass and spin) of the central source.

Meanwhile, high precision observational measurements of the electromagnetic spectrum reveals some unusual features from the known Kerr signals \cite[]{Johannsen-Psaltis2010a,Johannsen-Psaltis2010b,Johannsen-Psaltis2011}. Such discriminant is also observed in the gravitational waves spectrum from the BHs or neutron stars binary system \cite[]{abbott_2016observation, 2016PhRvL.116v1101A}. In these circumstances, several research groups have reported the parametric deviations to the Schwarzschild and Kerr black holes \cite[]{Johannsen_Psaltis_2011, Rezzolla_Zhidenko_2014, Konoplya_Zhidenko_2016detection,Konoplya-etal2016}. According to the no-hair theorem, such deviation to the original metrics brings alternative gravity theory ($i.e.$, the metrics are no longer the Einstein gravity solutions). Thus, we may anticipate that the non-Kerr spacetime can affect various strong gravity signatures and illustrate the peculiarities in the observations. In the last decade, one of the emerging and smeared alternative gravity is the Johannsen-Psaltis (JP) metric~\cite[]{Johannsen_Psaltis_2011}. They first include a deformation function, which contains infinite terms, in the Schwarzschild metric and then apply the Newman-Jenis algorithm to convert into a rotational Kerr-like metric. After that, deformation terms are restricted through the observational limitations on the weak-field modification of GR and asymptotic flatness. The finally obtained metric is characterized by the mass, spin and only one deviation parameter. When the deviation parameter is zero, it is reduced to the original Kerr metric. Their analysis also inflicts one valuable outcome in the calculation of the ISCO and circular photon orbits, and their dependency on the spin and deviation parameters under this proposed background. They show that, depending on the spin parameter, the central singularity of the spacetime becomes naked for outside observers when the deviation parameter crosses some limiting value. Usually, these irregularities in spacetime are described by the negative precession of the closed timelike orbits, which are the observational signature of the naked singular exotic objects.

Meanwhile, various investigations have been performed on the JP metric. For example, \citet[]{Bambi_2011} found the restriction to the spin parameter for non-Kerr BHs through the observational inconsistency in the radiative efficiency for luminous AGNs. In \citet[]{Bambi_2012}, the spacial topology of the event horizon for non-Kerr spacetime has been investigated. The properties of the ergosphere and the energy extraction by the Penrose process in a rotating deformed BH are carried out in \citet[]{liu_2012}. \citet[]{Chen_2012} analysed the strong gravitational lensing effect in a background of non-Kerr compact objects. \citet[]{Krawczynski_2012} distinguished between the original Kerr BHs hypothesis and non-Kerr BHs, and tested the no-hair theorem through the spectro-polarimetric observational data of the black hole XRBs. A detailed investigation of shadows and restriction to the spacetime parameters have been presented through the observation of polarization angels in \citet[]{Atamurotov_2013}. The inclusion of new parametric deviation approach and its challenge to the JP metric are encountered in \citet[]{Rezzolla_Zhidenko_2014}. A review on the signatures of alternative gravity by employing the gravitational waves from the BHs merging is presented in \citet[]{Yagi_Stein_2016}. A general ray-tracing formalism for black hole shadow calculations and its application to several deformed black holes have been reported in \citet[]{Younsi-etal2016}. The simultaneous existence of closed timelike orbits with negative precession and shadows is reported for the non-Kerr naked singular spacetime in \citet[]{Bambhaniya_et_al_2021}. Very recently, the properties of the accretion disc around a non-Kerr black hole without reflection symmetry have been revealed in \citet[]{Che-Yu-2021}. All these works evidently indicate that the JP metic attracts huge focus on it and also gets tremendous success for different applications in gravity. However, to the best of our knowledge, no one has conveyed the hydrodynamics of the accreting matter in the background of JP compact objects. Such deficiency in the literature pushes us to serve the present work, where we explore, for the very first time, the accretion dynamics of fluids in a spacetime of alternative gravity. We expect this analysis will lead to a better understanding of the non-Kerr spacetime in the light of accretion dynamics. 

In this work, we solve the general relativistic Euler's equation in the JP spacetime by utilizing the standard definitions of three velocities \cite[]{Lu1985} in a co-rotating frame. Even in the strong-field regime, flow equations mimics the Newtonian-like equations and provide the effective potential corresponding to the gravitating object \cite[]{Dihingia_et_al_2018}. We derive the radial velocity gradient and temperature gradient equations using the relativistic equation of state (REoS) that endure variable adiabatic index ($\Gamma$). After developing the mathematical framework, our primary motivation is to express the influence of the deformed term on the flow properties. We start our analysis accommodating the effect of the deformation parameter ($\varepsilon$) on the nature of critical points and the global transonic solutions around BH. Next, we separate the parameter space (in angular momentum ($\lambda$) - energy ($E$) plane) by means of the nature of the solution topologies and show their modifications with the input parameters. The global shock solutions, including their inherent properties, have been studied in detail. An important result is presented where we depict that even zero angular momentum flow can possesses multiple critical points and consequently suffer the shock transitions. This eventually provides new signatures of accretion dynamics in the non-Kerr BH spacetime. We show that  for a given spin parameter ($a_{\rm k}$), the usual BH accretion solutions continue to present up to a maximum value of the deformation parameter ($\varepsilon^{\rm max}$). When $\varepsilon > \varepsilon^{\rm max}$, the nature of the solution changes due to the presence of an extra critical point very close to the compact object. This possibly happens as the central source seems to appear as naked singularity \cite[]{Dihingia_et_al_2020} which is examined using numerical as well as analytical means. We further calculate a parameter space spanned by the spin ($a_{\rm k}$) and the deformation parameter ($\varepsilon$) according to the solution topologies around 
either BH or naked singularity state of the central objects. A comparison of $\varepsilon^{\rm max}$ obtained in the pseudo-Newtonian model and analytical approach is presented where good agreement is seen. This evidently
indicates that the accretion dynamics provides an alternative window to distinguish the subtle nature of the compact objects. 

Finally, we wish to emphasis that the global transonic accretion solutions in the deformed Kerr spacetime continue to exist as in the case of original Kerr spacetime. From our analysis, two new findings are imparted. One of them is the multiple critical point solutions including shock transitions. Another one is the existence of naked singularity for non-Kerr spacetime even if the spin parameter $a_{\rm k} < 1$. In the alternative gravity theory of GR, these specific findings can be considered observational evidence to distinguish the deformed Kerr spacetime from the original Kerr spacetime.

This paper is arranged as follows. In Section \ref{sec:assumptions_metric_governing_equations}, we develop the mathematical framework of the accretion disc theory and set up the critical point conditions. In Section \ref{sec:Hydrodynamics with deformation}, we present the effect of $\varepsilon$ on the critical point analysis, global flow solutions and modification of the parameter space for the non-Kerr BH. Section \ref{sec:Global accretion solution contain shock} analyses the shock-induced global accretion solutions and their parameter spaces. The dependence of shock properties on $\varepsilon$ is also established in this section. In Section \ref{sec:flow solutions associated with l = 0}, we show the flow solutions associated with zero angular momentum flows. In Section \ref{sec:naked_singularity}, we depict how $\varepsilon$ incorporates the naked singularity in the system through critical point analysis and their corresponding transonic solutions. In Section \ref{sec:a vs epsilon}, we represent the properties of the spacetime parameters in the JP metric. Finally, in Section \ref{sec:Conclusions}, we present conclusions.

\section{ASSUMPTIONS AND GOVERNING EQUATIONS}

\label{sec:assumptions_metric_governing_equations}

We present the basic equations governing the accretion flow using general relativistic hydrodynamics. To avoid mathematical complexity, the accretion disc is assumed to remain confine around the equatorial plane of the central object. We further consider the flow to be steady, inviscid and advective, where energy dissipations due to viscosity, thermal conduction, magnetic fields and radiative cooling are neglected.

\subsection{Governing equations}
\label{sec:hydrodynamics_deformed_Kerr} 

In the standard Boyer-Lindquist coordinates ($t, r, \theta, \phi$), the deformed Kerr metric (known as JP metric) is expressed as \cite[]{Johannsen_Psaltis_2011},
\begin{equation}
\label{eq:Kerr_spacetime}
\begin{split}
ds^{2} & = -(1-\frac{2M_{\rm BH}r}{\Sigma})\left[1+h(r,\theta)\right]dt^{2} \\& - \frac{4M_{\rm BH}r a_{\rm k}\sin^{2}\theta}{\Sigma}\left[1+h(r,\theta)\right]dtd\phi \\&+ \frac{\Sigma\left[1+h(r,\theta)\right]}{\Delta + a_{\rm k}^{2}h(r,\theta) \sin^{2}\theta} dr^{2} + \Sigma d\theta^{2} \\& +  \left[\Sigma + a_{\rm k}^{2}\left[1+h(r,\theta)\right](1+\frac{2M_{\rm BH}r}{\Sigma})\sin^{2}\theta\right]\sin^{2}\theta d\phi^{2}, 
\end{split}
\end{equation}
where $\Sigma = r^{2} + a_{\rm k}^{2}\cos^{2}\theta$ and $\Delta = r^{2} - 2M_{\rm BH}r + a_{\rm k}^{2}$. Here, $h(r,\theta) ~(= \varepsilon M_{\rm BH}^{3}r/\Sigma^{2})$ denotes the deformation that accounts the deviation of the metric under consideration from the original Kerr metric. And, $a_{\rm k}$ and $M_{\rm BH}$ are the spin parameter and the mass of the central object, respectively while $\varepsilon$ refers the deformation parameter. In $\varepsilon \rightarrow 0$ limit, equation (\ref{eq:Kerr_spacetime}) leads to the original Kerr metric. In this work, we use geometric units: $G = M_{\rm BH} = c = 1$, where $G$ and $c$ are the gravitational constant and the velocity of light, respectively.

In Introduction, we elaborately mention that the proposed form of the metric is inspired by various observational evidences. In particular, the suggested metric (equation (\ref{eq:Kerr_spacetime})) is put forward by considering the deviation of the usual Kerr solution, and the deformation parameter, measure of the spacetime deformation, is then constrained by the observational data (see e.g.~\cite[]{Johannsen_Psaltis_2011}). Because of this, the metric is not a solution of the vacuum Einstein's equations of motion. Under such circumstances, one is usually interested in searching for the gravity dynamics that predict such a solution. Accordingly, the present approach attempts to relate the notion of the alternate theory of gravity that differs from the Einstein's theory. Needless to mention that no such concrete theory of gravity has been investigated so far. Since the adopted spacetime introduces the deviations from the usual Kerr solution, and it has the potential to explain recent observational phenomena, there are growing interests to examine its influence in understanding the various astrophysical phenomena including its own physical properties. The accretion process around black hole is one of such phenomenon that is yet to be investigated for the black holes considering the above mentioned spacetime. Since the deformed Kerr metric offers novel features, the accretion dynamics are expected to be influenced, and hence, in the present work, we intend to investigate the properties of the accretion flow in the deformed spacetime of the rotating black hole.

In the general relativistic hydrodynamics, the mass conservation equation [$\nabla_k(\rho u^k) = 0$] and energy-momentum conservation equation [$\nabla_iT^{ik} = 0$] take the generalized form of the continuity and Euler's equations \cite[]{bahamonde-2015, Kumar_Chattopadhyay_2017, Dihingia_et_al_2018, Dihingia_Das_Nandi_2019, Dihingia_et_al_2020}, and are given by,
\begin{equation}
\nabla_{k}\left[(e+p)u^{k}\right] = u^{k}\nabla_{k}p
\label{eq:continuity_eq}
\end{equation}
and
\begin{equation}
(e+p)u^{k}\nabla_{k}u_{i} + \nabla_{i}p + u_{i}u^{k}\nabla_{k}p = 0.
\label{eq:Euler's_eq}
\end{equation}
Here $e$, $p$ and $u^{k}$ are the total internal energy, pressure, and four-velocity of the perfect fluid, respectively, and the spacetime indices $(i, k)$ bear values from $0$ to $3$.

We consider the fluid to obey the same symmetries of the spacetime, and apply the condition $\xi^{\mu}\nabla_{\mu}\mathcal{Q} = 0$, where $\mathcal{Q}$ refers any fluid parameters ($e. g.$, mass density, pressure and four-velocity, etc.) and $ \xi^{\mu}$ is a generic Killing vector associated with the spacetime. Therefore, using equations (\ref{eq:continuity_eq}) and (\ref{eq:Euler's_eq}), we get the conservation equation as~\cite[]{Ahmed-2016, azreg-2017, Dihingia_et_al_2018, azreg-2018, Dihingia_et_al_2020},

\begin{equation}
	\label{eq:lie3}
	u^{\nu}\nabla_{\nu}(\mathfrak{h}u_{\mu}\xi^{\mu}) = 0.
\end{equation}

The stationary and axisymmetric spacetime, due to its symmetries, is associated with two Killing vectors: $\eta^{i} = \delta^{i}_{t}$ and $\zeta^{i} = \delta^{i}_{\phi}$. The corresponding conserved quantities are given by,
\begin{equation}
\label{eq:killing_conservation}
-\mathfrak{h}u_{t} = E \hspace{0.5cm} \text{and} \hspace{0.5cm} \mathfrak{h}u_{\phi} = \mathcal{L},
\end{equation}
where $\mathfrak{h} = (e+p)/\rho$ is the specific enthalpy and $\rho$ is the mass density of the fluid. Here, $E$ is the Bernoulli function and $ \mathcal{L}$ is the bulk angular momentum per unit mass of the fluid. The specific angular momentum is defined as $\lambda = \mathcal{L}/E = - u_{\phi}/u_{t}$ that remains conserved along the streamlines of the flow (see Eq.~(\ref{eq:killing_conservation})).

Following \citet[]{Lu1985}, we adopt the components of three velocity in a co-rotating frame. The azimuthal, polar and radial component of three-velocity are defined as 
$ v_{\phi}^{2} = u^{\phi}u_{\phi}/(-u^{t}u_{t})$, $v_{\theta}^{2} = \gamma_{\phi}^{2}(u^{\theta}u_{\theta})/(-u^{t}u_{t})$ and $v^{2} = \gamma_{\phi}^{2}\gamma_{\theta}^{2}v_{r}^{2}$, respectively, where $v_{r}^{2} = u^{r}u_{r}/(-u^{t}u_{t})$.
The respective bulk Lorentz-factors ($\gamma_{\phi}$, $\gamma_{\theta}$ and $\gamma_{v}$) are expressed as $\gamma_{\phi}^{2} = 1/(1-v_{\phi}^{2})$, $\gamma_{\theta}^{2} = 1/(1-v_{\theta}^{2})$ and $\gamma_{v}^{2} = 1/(1-v^{2})$. We further consider the fluid motions around the disc equatorial plane ($\theta = \pi/2$) with $v_{\theta} = 0$ and $\gamma_{\theta} = 1$. Under these assumptions, the angular velocity of the fluid is obtained as \cite[]{Chakrabarti_1996},
\begin{equation}
\label{eq:omega1}
\Omega = \frac{u^{\phi}}{u^{t}} = \frac{\left[\lambda (r-2) + 2a_{k}\right](1+h)}{a_{k}^{2}(1+h)(r+2)-2a_{k}\lambda (1+h) +r^{3}},
\end{equation}
where $h = h(r, \theta = \pi/2) = \varepsilon/r^{3}$. The normalization $u^{i}u_{i} = -1$ yields the covariant time-component of four-velocity and is given by \cite[]{Dihingia_Das_Nandi_2019, Dihingia_et_al_2020},
\begin{equation}
\label{eq:u_t}
\begin{split}
& u_{t} =  \gamma_{v} \\ & \times \sqrt{\frac{(\Delta+a_{k}^{2}h)(1+h)r}{a_{k}^{2}(r+2)(1+h) - 4a_{k}\lambda(1+h) + r^{3} -\lambda^{2}(r-2)(1+h) }}.
\end{split}
\end{equation}
The above relation is explicitly written for our metric (\ref{eq:Kerr_spacetime}). Since the motion is considered around the disk equatorial plane, and there are time translation as well as azimuthal symmetries, the Eqs.~(\ref{eq:continuity_eq}) and (\ref{eq:Euler's_eq}) for $i=r$ are simplified as \cite[]{ Dihingia_et_al_2018, Dihingia_et_al_2020},
\begin{equation}
\label{eq:entropy_generation_1}
\frac{e+p}{\rho}\frac{d\rho}{dr} - \frac{de}{dr} = 0
\end{equation}
and 
\begin{equation}
\label{eq:mimic1}
\gamma_{v}^{2}v\frac{dv}{dr} + \frac{1}{\mathfrak{ h}\rho}\frac{dp}{dr} + \frac{d\Phi^\text{eff}}{dr} = 0.
\end{equation}
For the same reason, $i=\theta$ component of Eq.~(\ref{eq:Euler's_eq}) is trivially satisfied and therefore does not lead to any new equation. However, the equations for both $i = t~{\rm and}~\phi$ govern the conservation of specific angular momentum \cite[]{Dihingia_et_al_2018}, which we already encountered through the Killing symmetries in the metric. Following \cite[][and references therein]{Chakrabarti_1989,Dihingia_et_al_2018}, we identify $\Phi^\text{eff}$ in Eq.~(\ref{eq:mimic1}) as the effective pseudo-potential and is given by ,
\begin{equation}
\label{eq:Phi_eff}
\Phi^\text{eff} = 1 + 0.5{\ln}(\Phi)~.
\end{equation}
For the metric given in Eq.~(\ref{eq:Kerr_spacetime}),  $\Phi$ is given by
\begin{equation}
\label{eq:Phi}
\Phi = \frac{(\Delta+a_{\rm k}^{2}h)(1+h)r}{a_{\rm k}^{2}(r+2)(1+h) - 4a_{\rm k}\lambda(1+h) + r^{3} -\lambda^{2}(r-2)(1+h)}.
\end{equation}

Integrating the mass conservation equation, we obtain the globally conserved mass accretion rate ($\dot{M}$) which is given by \cite[]{Kumar_Chattopadhyay_2017,Dihingia_et_al_2018}, 
\begin{equation}
\label{eq:mass_accretion}
\dot{M} = -4\pi \rho v \gamma_{v} H\sqrt{(\Delta + a_{\rm k}^{2}h)(1+h)},
\end{equation}
where $H$ is the local half thickness of the disc. Following the work of 
\citet[]{Ryu2009}, we adopt the relativistic equation of state (REoS) and pressure ($p$) as,
\begin{equation}
\label{eq:reos}
\hspace{2.3cm}  e = \frac{\rho f}{\tau} \hspace{0.3cm} \text{and} \hspace{0.3cm} p = \frac{2\rho\Theta}{\tau}, 
\end{equation}
where $\tau = 2 - \xi(1- 1/\chi)$, the composition ratio $\xi = n_{p}/n_{e}$ and the mass ratio $\chi = m_{e}/m_{p}$. The number density and the mass of the $i$th species (electron, proton) are denoted by $n_{i}\in$ $\{n_{e}, n_{p}\}$ and $m_{i}\in$ $\{m_{e}, m_{p}\}$, respectively. Moreover, we consider $\xi = 1$, throughout our analysis. Here the quantity $f$ is obtained in terms of dimensionless temperature ($\Theta = k_{B}T/m_{e}c^{2}$, $k_B$ is the Boltzmann constant and $T$ is the flow temperature in Kelvin) as 
\begin{equation}
\label{eq:f}
f = (2 - \xi)\left[1 + \Theta \left(\frac{9\Theta + 3}{3\Theta + 2}\right)\right] + \xi\left[\frac{1}{\chi} + \Theta \left(\frac{9\Theta + 3/\chi}{3\Theta + 2/\chi}\right)\right].
\end{equation}
For REoS, polytropic index ($N$), adiabatic index ($\Gamma$) and sound speed ($C_{s}$) are defined as
\begin{equation}
\label{eq:polytropic_index and sound_speed}
N = \frac{1}{2}\frac{df}{d\Theta}; \hspace{0.2cm} \Gamma = 1 + \frac{1}{N}; \hspace{0.2cm} \text{and} \hspace{0.2cm} C_{s}^{2} = \frac{\Gamma p}{e+p} = \frac{2\Gamma \Theta}{f + 2\Theta}.
\end{equation}

Considering the hydrodynamic equilibrium in the vertical direction, the local half thickness of the disc ($H$) is calculated as \cite[]{Lasota_1994, Riffert1995, Peitz1997},
\begin{equation}
\label{eq:hafl_thickness}
H = \sqrt{\frac{pr^{3}}{\rho F}} = \sqrt{\frac{2r^{3}\Theta}{\tau F}},
\end{equation}
where 
\begin{equation}
\label{eq:F}
F = \gamma_{\phi}^{2} \frac{(r^{2} + a_{\rm k}^{2})^{2} + 2\Delta a_{\rm k}^{2}}{(r^{2} + a_{\rm k}^{2})^{2} - 2\Delta a_{\rm k}^{2}}.
\end{equation}
Integrating Eq.~(\ref{eq:entropy_generation_1}) and using  Eq.~(\ref{eq:reos}), the mass density is obtained as \cite[]{Kumar_et_al_2013, Chattopadhyay_Kumar_2013, Dihingia_et_al_2020},
\begin{equation}
\label{eq:dendity_temp_dependence}
\rho = \mathcal{K}\exp{(k_{3})}\Theta^{3/2}(3\Theta + 2)^{k_{1}}(3\Theta+2/\chi)^{k_{2}}, 
\end{equation}
where $k_{1} = 3(2-\xi)/4$, $k_{2}=3\xi/4$, $k_{3}=(f-\tau)/(2\Theta)$, and $\mathcal{K}$ refers the entropy constant. Using Eqs.~(\ref{eq:mass_accretion}) and (\ref{eq:dendity_temp_dependence}), we compute the entropy accretion rate as \cite[]{Chattopadhyay_Kumar_2016, Kumar_Chattopadhyay_2017},
\begin{equation}
\label{eq:entropy_accretion_rate}
\begin{split}
\mathcal{\dot{M}} = \frac{\dot{ M}}{4\pi\mathcal{K}} & = v\gamma_{v}H\sqrt{(\Delta+a_{k}^{2}h)(1+h)}\\ & \times \exp{(k_{3})}\Theta^{3/2}(3\Theta + 2)^{k_{1}}(3\Theta+2/\chi)^{k_{2}}.
\end{split}
\end{equation}
Considering logarithmic derivative of Eq.~(\ref{eq:mass_accretion}) and setting the condition of constant mass accretion rate ($i.e.$, $d{\dot{ M}}/dr = 0$), the temperature gradient is expressed as,
\begin{equation}
\label{eq:temperature-gradient}
\begin{split}
& \frac{d\Theta}{dr} = -\frac{2\Theta}{2N + 1} \\ & \times \left[  \frac{\gamma_{v}^{2}}{v}\frac{dv}{dr} + N_{11} + N_{12} - \frac{3\varepsilon}{2r^{4}}\left( \frac{a_{\rm k}^{2}}{\Delta + a_{\rm k}^{2}h} + \frac{1}{1+h}\right) \right]~,
\end{split}
\end{equation}
where 
\begin{equation}
\label{eq:N_11}
N_{11} = \frac{5}{2r} + \frac{r-a_{\rm k}^{2}(1+h)}{r(\Delta + a_{\rm k}^{2}h)} \hspace{0.2cm} \text{and} \hspace{0.2cm} N_{12} = -\frac{1}{2F}\frac{dF}{dr}.
\end{equation}
The explicit form of the quantity $\frac{1}{F}\frac{dF}{dr}$ is obtained by taking the logarithmic derivative of Eq.~(\ref{eq:F}) and is given by,
\begin{equation}
\label{log_diff_F}
\frac{1}{F}\frac{dF}{dr} = \gamma_{\phi}^{2}\lambda \Omega^{'} + 4a_{\rm k}^{2}(r^{2} + a_{\rm k}^{2}) \left[  \frac{(r^{2}+a_{\rm k}^{2})\Delta^{'} - 4\Delta r}{(r^{2}+a_{\rm k}^{2})^{4} - 4\Delta^{2}a_{\rm k}^{4}} \right],
\end{equation} 
\begin{widetext}
where
\begin{equation}
\label{eq:Delta_prime}
\begin{split}
\Delta^{'} = \frac{d\Delta}{dr} = 2 (r-1); \hspace{0.2cm}\Omega^{'} = \frac{d\Omega}{dr} = & -2(1+h) \left[ \frac{a_{\rm k}^{3}(1+h)-2a_{\rm k}^{2}\lambda (1+h) + a_{\rm k}[\lambda^{2}(1+h)+3r^{2}] + r^{2}\lambda (r-3)}{\left[a_{\rm k}^{2}(r+2)(1+h) - 2a_{\rm k}\lambda (1+h) + r^{3}\right]^{2}} \right] \\ & - \frac{3\varepsilon\left[\lambda (r-2) + 2a_{\rm k}\right]}{r\left[ a_{\rm k}^{2}(r+2)(1+h) - 2a_{\rm k}\lambda (1+h) + r^{3}\right]^{2}}.
\end{split}
\end{equation}
\end{widetext}

Finally, we capitalize Eq.~(\ref{eq:polytropic_index and sound_speed}) and obtain the radial velocity gradient from Eq.~(\ref{eq:mimic1}) as,
\begin{equation}
\label{eq:wind}
\frac{dv}{dr} = \frac{{\mathcal N}}{{\mathcal D}},
\end{equation} 
where the explicit form of the denominator ${\mathcal D}$ and that of the numerator ${\mathcal N}$ are represented by,
\begin{equation}
\label{eq:D}
{{\mathcal D}} = \gamma_{v}^{2}\left[v - \frac{2C_{s}^{2}}{(\Gamma + 1)v}\right]
\end{equation}
and
\begin{equation}
\label{eq:N}
\begin{split}
 {\mathcal N} = & \frac{2C_{s}^{2}}{\Gamma + 1} \\ \times & \left[  N_{11} + N_{12} - \frac{3\varepsilon}{2r^{4}}\left( \frac{a_{\rm k}^{2}}{\Delta + a_{\rm k}^{2}h} + \frac{1}{1+h}\right)\right] - \frac{d\Phi^\text{eff}}{dr}.
\end{split}
\end{equation}

\subsection{Critical point conditions}
\label{sec:critical_point_condition} 

In an accretion process around a gravitating object, infalling matter starts accreting with negligible radial velocity from the outer edge of the disk (usually far away from the horizon) and remain sub-sonic. On the other hand, accretion flow enters into the black hole super-sonically in order to satisfy the inner boundary conditions imposed by the event horizon. Since the motion of the flow generally remains smooth everywhere, accreting matter experiences sonic state transition at some point to become transonic \cite[]{Liang1980,Abramowicz1981} and such a point is referred as critical point ($r_c$). At $r_c$, the radial velocity gradient takes $(dv/dr)_{r_c}=0/0$ (Eq.~\ref{eq:wind}) form as it must be real and finite, and hence, we obtain the critical point conditions by setting ${\mathcal D}={\mathcal N}=0$ simultaneously which are given by,

\begin{equation}
\label{eq:D=0}
v_{c}^{2} = \frac{2C_{sc}^{2}}{(\Gamma_{c} + 1)}
\end{equation}
and 

\begin{equation}
\label{eq:N=0}
\begin{split}
C_{sc}^{2} &= \frac{\Gamma_{c} + 1}{2}\left(\frac{d\Phi^\text{eff}}{dr}\right)_{c} \times \\&  
\left[ (N_{11})_{c} + (N_{12})_{c} - \frac{3\varepsilon}{2r_{c}^{4}}\left( \frac{a_{\rm k}^{2}}{\Delta_{c} + a_{\rm k}^{2}h_{c}} + \frac{1}{1+h_{c}}\right)\right]^{-1},
\end{split}
\end{equation}
where the subscript \say{c} refers quantities measured at the critical point. We evaluate $(dv/dr)_{r_c}$ by applying l$'$H\^{o}pital's rule which is obtained as
\begin{equation}
\label{eq:dv/dr_r_c}
\frac{dv}{dr}\bigg\vert_{c} = \frac{-B\pm\sqrt{B^{2}-4AC}}{2A},
\end{equation} 
where the resulted form of quantities $A$, $B$ and $C$ are presented in Appendix \ref{appe:appendix}. In general, critical points are classified in three different categories. For {\it saddle} type critical points, both values of $(dv/dr)_{r_{c}}$ are real with opposite sign. For {\it nodal} type critical points, both values of $(dv/dr)_{r_{c}}$ are real and same sign, whereas for O-type critical point, $(dv/dr)_{r_{c}}$ becomes imaginary. It is noteworthy that any physically acceptable accretion solution only passes through the saddle type critical point \cite[and references therein]{Das_2007, Chakrabarti_Das_2004}, and hence, in this work, we focus only those accretion solutions that possess saddle type critical point (hereafter critical point). We further mention that accretion flow may contain multiple critical points depending of the flow parameters and flow of this kind are potentially favourable to contain shock wave (see Section \ref{sec:Global accretion solution contain shock}).

\section{Hydrodynamics with deformation}
\label{sec:Hydrodynamics with deformation}

In this section, we explore the role of the deformation parameter ($\varepsilon$) in deciding the nature of the critical points as well as the accretion solutions in deformed Kerr spacetime. While doing this, we identify the range of parameters that allows accretion solutions around black holes. We also put efforts in examining the nature of the accretion solutions beyond black hole environment as well.

\subsection{Critical points analysis}

\label{sec:Critical_point_analysis}

\begin{figure}
	\centering
	\includegraphics[width=\columnwidth]{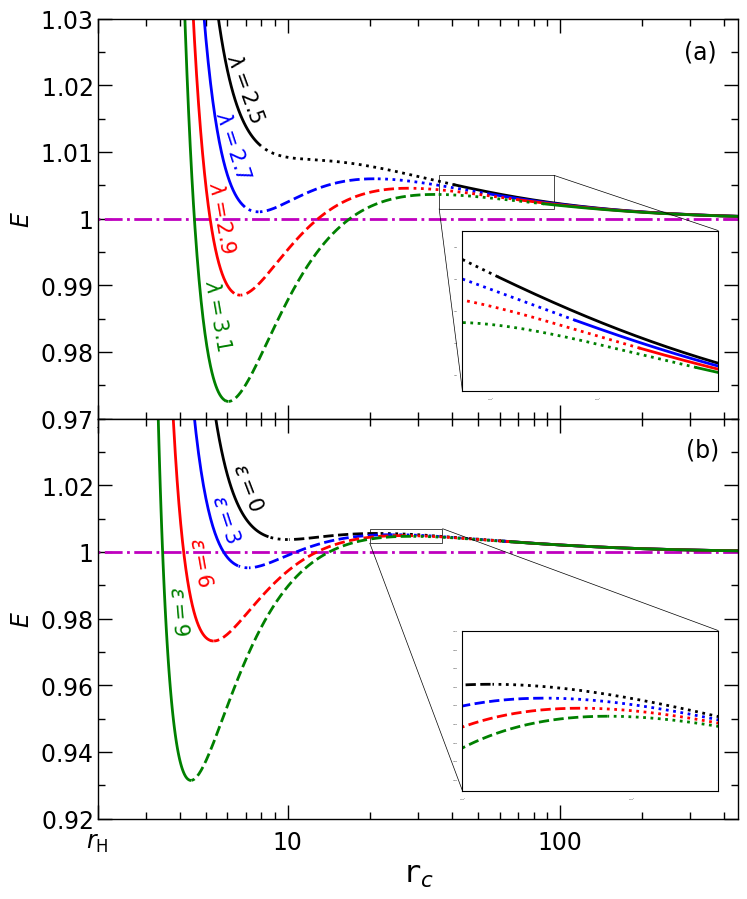}
	\caption{Plot of specific energy ($E$) as the function of critical point locations ($r_{c}$) for (a) different angular momentums $\lambda$ = 2.5 (black), 2.7 (blue), 2.9 (red) and 3.1 (green) with deformation parameter $\varepsilon = 3$, and (b) different deformation parameters $\varepsilon$ = 0 (black), 3 (blue), 6 (red) and 9 (green) with the specific angular momentum $\lambda = 2.8$. Solid, dotted and dashed curves denote saddle, nodal and O-type critical points, respectively. The dot-dashed horizontal line indicates the specific energy $E=1$. In each panel, we zoom a part of the plots for the purpose of clarity. See text for details.
	} 
	\label{fig:E_vs_R_different_l_and_epsilon}
\end{figure}

As the accretion solutions embrace the critical points, we start our analysis by understanding the nature of critical points. For that we calculate the specific energy ($E$) at a critical point ($r_{c}$) by solving equations (\ref{eq:killing_conservation}), (\ref{eq:D=0}) and (\ref{eq:N=0}) using the global parameters, namely $\lambda$, $\varepsilon$ and $a_{\rm k}$, respectively. In Fig.~\ref{fig:E_vs_R_different_l_and_epsilon}, the variation of $E$ as a function of $r_{c}$ is presented for different $\lambda$ with $\varepsilon = 3$ (see panel (a)) and for different $\varepsilon$ with $\lambda = 2.8$ (see panel (b)). Presently, we choose the Kerr parameter $a_{\rm k}$ = 0. However, we mention that there are no qualitative differences between the characteristics of critical points for the non-spinning and spinning black holes. Hence, in this work, most of the analyses have been carried out considering $a_{\rm k} = 0$, although there are instances where results for $a_{\rm k}\neq 0$ are presented according to the necessity. In the figure, different $\lambda$ and $\varepsilon$ values are marked in the respective panels. We use black, blue, red and green curves for $\lambda$ = 2.5, 2.7, 2.9 and 3.1 respectively. The same color sequence is used to represent results for $\varepsilon$ = 0, 3, 6, and 9, respectively. In each panel, a given curve is generally comprised with saddle, nodal and O-type critical points and they are demonstrated by the solid, dotted and dashed curves respectively. Moreover, these critical points appear in sequence as saddle-nodal-spiral-nodal-saddle as $r_c$ is increased. In addition, we observe that all curves have an asymptotic behavior towards $E \simeq 1$ (dot-dashed horizontal line) for larger values of $r_c$ irrespective of $\lambda$ and $\varepsilon$ values. Depending on $E$, $\lambda$ and $\varepsilon$, flow may contain either single or multiple critical points. Usually, critical points formed near and far away from the the horizon are called as inner ($r_\text{{in}}$) and outer critical points ($r_\text{{out}}$), respectively. It is evident from the figure that there exists a range of $E$ that yields multiple critical points and such energy range is strictly depends on $\lambda$ and $\varepsilon$ values. Following this, in Section \ref{sec:parameter space}, we put effort to identify the effective region of the parameter space based on the nature of the accretion solutions. Overall, it is now evident that $\varepsilon$, $\lambda$ and $E$ play pivotal role in determining the nature of the critical points and its associated properties. 

\subsection{Effect of ${\varepsilon}$ on the global accretion solutions}

\label{sec:Global accretion solution}

\begin{figure*}
	\includegraphics[width=\linewidth]{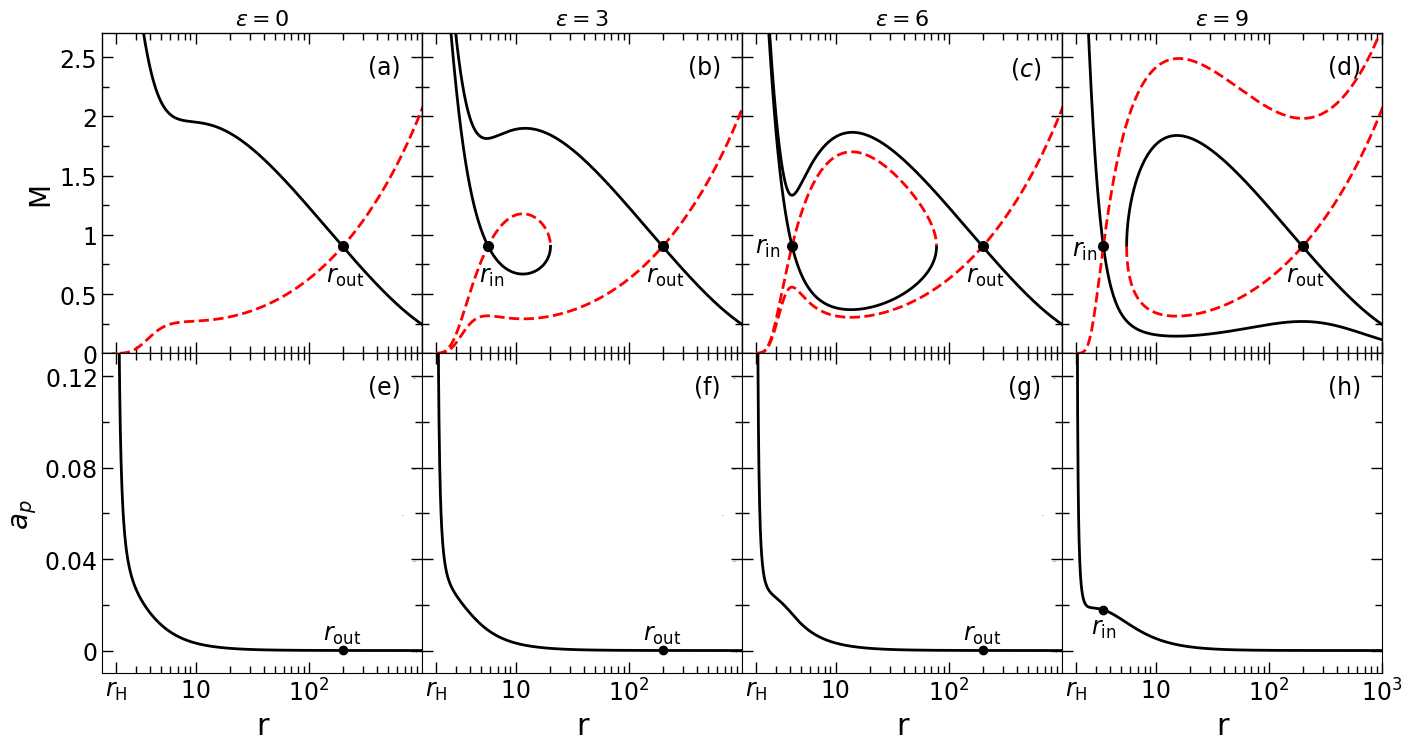}
	\caption{{\it Top panels:} Plot of Mach number $M$ ($= v/C_{s}$) as a function of radial distance $r$. Solid and dashed curves denote the accretion solution and wind solution, respectively. Filled circles denote the critical points. In this figure, we choose $a_{k} = 0$, $\lambda = 2.8$ and $E = 1.001$. Results in panels (a), (b), (c) and (d) are obtained for $\varepsilon = 0, 3, 6$ and $9$, respectively. {\it Bottom panels:} Plot of the magnitude of four-acceleration ($\rm {a}_{p}$) with $r$ corresponding to the global accretion solution (solid) presented in the top panel. See text for details.
	}
	\label{fig:M_vs_r_different_epsilon}
\end{figure*}

Here, we examine the impact of $\varepsilon$ on the accretion solutions. While doing so, we calculate the location of the critical point ($r_{c}$), and the corresponding radial velocity ($v_{c}$) and dimensionless temperature ($\Theta_{c}$) at $r_c$ by simultaneously solving Eqs. (\ref{eq:killing_conservation}), (\ref{eq:polytropic_index and sound_speed}), (\ref{eq:D=0}) and (\ref{eq:N=0}) for a given set of input parameters ($\lambda$, $E$, $a_{k}$ and $\varepsilon$). We employ $\Theta_{c}$ and $v_{c}$ as the initial values at $r_c$ to simultaneously solve Eqs. (\ref{eq:temperature-gradient}) and (\ref{eq:wind}) once inward up to the horizon ($r_{\text{H}}$) and then outward up to the outer edge of the disk ($r_\text{{edge}}$). Finally, we join this two parts of the solution to obtain the complete radial profiles of velocity ($v$) and temperature ($\Theta$). In Fig. \ref{fig:M_vs_r_different_epsilon}, we depict the accretion solutions ($M$ vs. $r$) for different $\varepsilon$, where $E = 1.001$, $\lambda = 2.8$ and $a_{\rm k} = 0$ are chosen. In panels (a-d), the variation of Mach number ($M$) as function of radial distances ($r$) is presented for $\varepsilon = 0, 3, 6$ and $9$, respectively. Here, the solid curve represents the accretion solution whereas dashed curve denotes the corresponding wind solution. In the figure, filled circles refer the critical points, where inner ($r_\text{in}$) and outer 
($r_\text{out}$) critical points are marked. We observe that for $\varepsilon = 0$, the flow passes through the outer critical point at $r_{\text{out}} = 198.6333$ and connects the outer edge of the accretion disc ($r_{\text{edge}} = 1000$) to the black hole horizon ($r_{\text{H}}$) (see panel (a)). Solutions of this kinds are often called  by some authors as {\it heteroclinic} solutions \cite[]{Ahmed-2015, mustapha-2017}. As the deformation parameter is increased (say $\varepsilon = 3$) keeping other input parameters unchanged, inner critical point is appeared at $r_{\text{in}} = 5.7136$ along with the outer critical point at $r_{\text{out}} = 198.5498$. Interestingly, the solution passing through the outer critical point continues to connect $r_{\text{H}}$ and $r_{\text{edge}}$, however the inner critical point solutions fail to do so as it terminates at a radius $r_{\rm t}=20.3336$ in between inner and outer critical points as $r_{\text {in}} < r_t < r_{\text {out}}$  as shown in panel (b). The closed transonic solutions that unable to connect $r_{\rm edge}$ and $r_{\rm H}$ are also called as {\it homoclinic} solutions ~\cite[]{Ahmed-2016,Ahmed-2015}. For $\varepsilon = 6$, the nature of the flow solution remains qualitatively similar to panel (b) although $r_{\rm t}=77.7947$ is increased (see panel (c)). We wish to emphasize that solutions presented in panel (b-c) may experience shock transition and we plan to discuss it elaborately in Section \ref{sec:Global accretion solution contain shock}. With the further increase of deformation parameter $\varepsilon = 9$, the solution characteristics are changed completely as shown in panel (d). We find that the solution passing through $r_{\text{in}} = 3.4628$ smoothly connects $r_{\text{edge}}$ to $r_{\text{H}}$, but the possesses $r_{\text{out}} = 198.3757$ fails to do so. Hence, it is evident that $\varepsilon$ plays a decisive role in determining the nature of the accretion solutions around the central object under consideration. In Fig. \ref{fig:M_vs_r_different_epsilon}e-h, we present the variation of the magnitude of proper four-acceleration $\rm {a}_{p}$ ($= \sqrt{{\rm a}^{i}{\rm a}_{i}}$, where four-acceleration $ {{\rm a}^{i}} = u^{k}\nabla_{k}u^{i}$ \cite[]{azreg-2017}) with $r$ corresponding to the global transonic solutions depicted in panels (a), (b) (c) and (d), respectively. In each panel, we notice that $\rm {a}_{p}$ increases as flow accretes towards the horizon and diverges at $r_{\rm H}$ as $v \rightarrow 1$. The divergent nature of $\rm {a}_{p}$ at the horizon is quite consistent with the theoretical prediction; e.g. see discussion in section $6.3$ of \cite{Carroll}. This evidently indicates that, with respect to any observer at static infinity, instead of timelike fluid only photons can stay at $r = r_{\rm H}$ . In a way, this defines a boundary region of the spacetime (usually called static limit) of the black hole, where $g_{\rm tt}(r_{\rm H}) = 0$. 

\subsection{Parameter space based on nature of accretion solutions}

\label{sec:parameter space} 

\begin{figure}
	\begin{center}
		\includegraphics[width=\columnwidth]{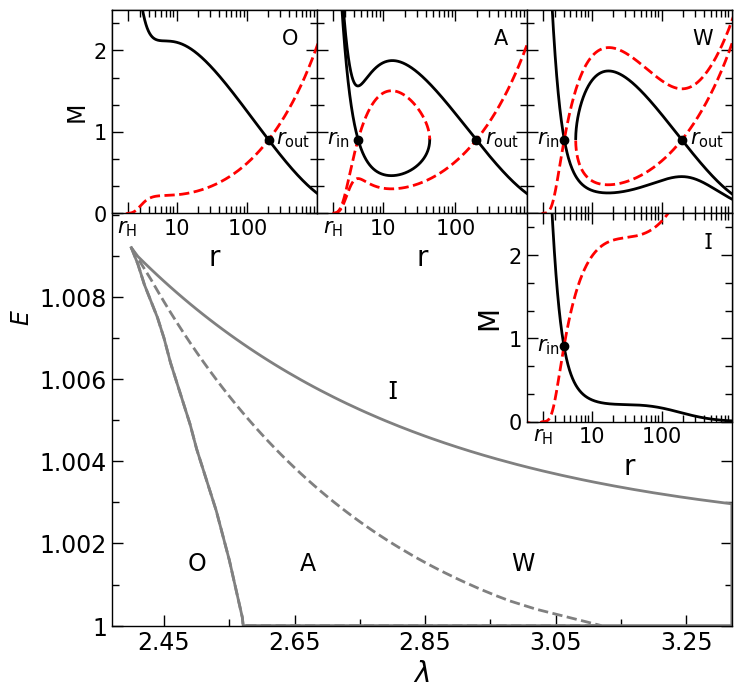}
	\end{center}
	\caption{Division of parameter space in $\lambda-E$ plane according to the nature of the flow solutions. Here, we fix $\varepsilon = 5$. Four regions are identified which are marked as O, A, W and I. Examples of representative flow solutions from individual regions are shown in each panel. See text for details.		
	}
	\label{fig:division_parameter_space}
\end{figure} 

In this section, we separate the effective region of the parameter space in $\lambda - E$ plane according to the nature of the accretion solutions. The obtained results are plotted in Fig. \ref{fig:division_parameter_space}, where parameter space is divided in to four regions marked as O, A, W and I. Examples of different solution topologies obtained using flow parameters ($\lambda, {\cal E}$) from these regions are depicted at the insets where Mach number ($M = v/C_{s}$) is plotted with the radial distances ($r$). Here, we choose $a_{\rm k}=0$ and $\varepsilon = 5$. In each panel, the accretion and wind solutions are presented using solid (black) and dashed (red) curves and the filled circle denotes the critical point. Inside the bounded region of the parameter space, flow solutions are found to possess multiple critical points which is further sub-divided based on the entropy accretion rate (${\dot{\mathcal M}}$) measured at the critical points. Accordingly, region A and B are obtained for ${\dot{\mathcal M}}(r_{\rm in}) > {\dot{\mathcal M}}(r_{\rm out})$ and ${\dot{\mathcal M}}(r_{\rm in}) < {\dot{\mathcal M}}(r_{\rm out})$, respectively. For flow solutions presented in panel O, we choose ($\lambda, E$) $= (2.5, 1.001)$ and get only outer critical points at $r_\text{out} = 203.562$. We set ($\lambda, E$) $= (2.80, 1.001)$ to obtain the flow solutions in panel A, where $r_{\text{in}} = 4.5132$ and $r_{\text{out}} = 198.4749$. Similarly, for panel W, we fix ($\lambda, E$) $= (3.0, 1.007)$ and find $r_{\text{in}} = 4.0795$ and $r_{\text{out}} = 194.5943$. Finally, in panel I, we choose ($\lambda, E$) $= (3.0, 1.0045)$ that yields only inner critical point at $r_{\text{in}} = 4.0445$. Note that all the above findings are in general qualitatively similar to the results obtained for pure Kerr black hole \cite[]{Dihingia_et_al_2018}, but differs quantitatively due to the deformation present in the adopted spacetime. The nature of the accretion flows obtained from the different regions in Fig. \ref{fig:division_parameter_space} is summarized in Table \ref{tab:table-1}.

\begin{table}
	\centering
	\caption{The nature of the accretion solutions presented in Fig.~\ref{fig:division_parameter_space}.} 
	\label{tab:table-1}
	\begin{ruledtabular}
		\begin{tabular}{ll}
			 Type & Nature of accretion solutions\\
			\hline
		     O & Open solution containing $r_{\rm out}$\\ \\
			 A & Closed solution containing $r_{\rm in}$ and \\
			  & open solution containing $r_{\rm out}$ \\ \\
			 W & Open solution containing $r_{\rm in}$ and \\
			& closed solution containing $r_{\rm out}$ \\ \\ 
			 I & Open solution containing $r_{\rm in}$\\ 
		\end{tabular}
	\end{ruledtabular}
\end{table} 

\begin{figure}
	\begin{center}
		\includegraphics[width=\columnwidth]{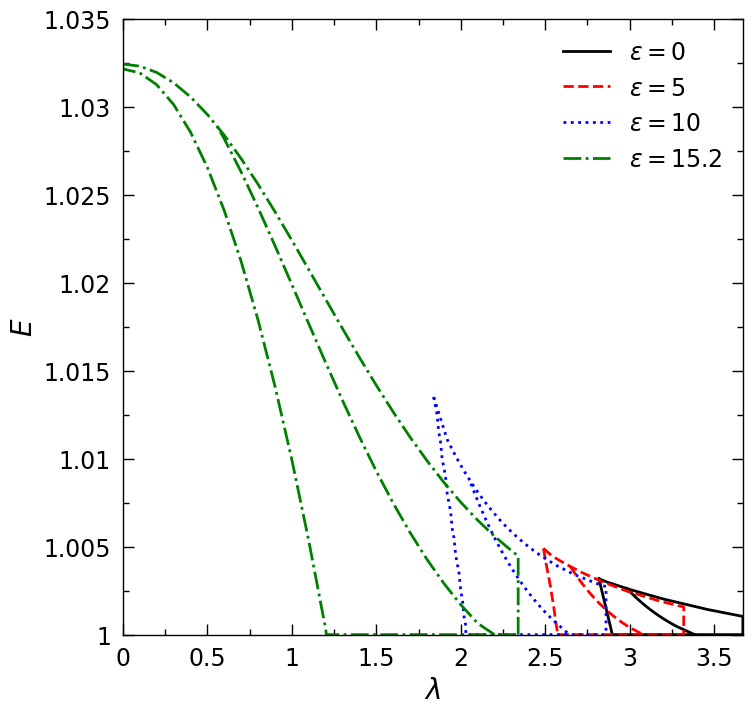}
	\end{center}
	\caption{Modification of the parameter space (in $\lambda-E$ plane) for multiple critical points due to the increase of the deformation parameter ($\varepsilon$). Regions bounded with solid (black), dashed (red), dotted (blue) and dot-dashed (green) curves are for $\varepsilon = 0, 5, 10$ and $15.2$, respectively. In each parameter space, middle curve refers $\dot{\mathcal M}(r_{\rm in})=\dot{\mathcal M}(r_{\rm out})$. See text for details.
	}
	\label{fig:comparison_parameter_space_fixed_a}
\end{figure} 

Next, we examine the modification of the parameter space for multiple critical points in $\lambda-{\cal E}$ plane for different $\varepsilon$. We present the obtained results in Fig. \ref{fig:comparison_parameter_space_fixed_a}, where $a_{\rm k} = 0$ is used, and the regions bounded by solid (black), dashed (red), dotted (blue) and dot-dashed (green) are for $\varepsilon = 0, 5, 10$ and $15.2$, respectively. Each parameter space is further sub-divided based on the entropy accretion rate at the critical points, where middle curve denotes $\dot{\mathcal M}(r_{\rm in})=\dot{\mathcal M}(r_{\rm out})$. We observe that the effective domain of the parameter space is increased and also shifted towards lower angular momentum and higher energy sides as $\varepsilon$ is increased. Further, for the first time to the best of our knowledge, we report that for $\varepsilon = 15.2$, multiple critical points continue to exist even for zero angular momentum flow. Indeed, the choice of $\varepsilon = 15.2$ is not arbitrary, in fact, it is the lower limit of the deformation parameter that provides multiple critical points for $\lambda = 0$. Needless to mention that such a lower limit is not universal, instead it depends on the spin parameter ($a_{\rm k}$).

\begin{figure}
	\includegraphics[width=\columnwidth]{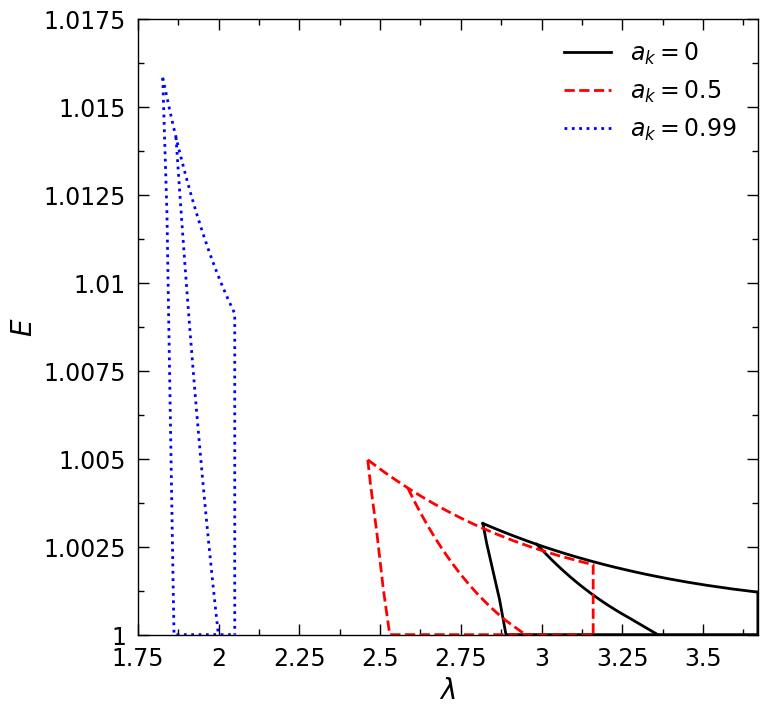}
	\caption{Modification of the parameter space for multiple critical points for different Kerr parameter ($a_{\rm k}$). The area within solid (black), dashed (red) and dotted (blue) boundaries are for $a_{\rm k} = 0, 0.5$ and $0.99$, where $\varepsilon=0.02$ is chosen. In each parameter space, middle curve refers $\dot{\mathcal M}(r_{\rm in})=\dot{\mathcal M}(r_{\rm out})$. See text for details. 
	}
	\label{fig:comparison_parameter_space_fixed_epsilon}
\end{figure}

Fig. \ref{fig:comparison_parameter_space_fixed_epsilon} shows the alteration of the parameter space for multiple critical points for different Kerr parameters ($a_{\rm k}$). Here, we choose $\varepsilon = 0.02$. The region bounded using solid (black), dashed (red) and dotted (blue) curves correspond to the result for $a_{\rm k}= 0, 0.5$ and $0.99$ respectively. In each parameter space, middle curve refers $\dot{\mathcal M}(r_{\rm in})=\dot{\mathcal M}(r_{\rm out})$. It is evident from the figure that parameter space is moved to the lower angular momentum and higher energy domains with the increase of the black hole spin. With this, we indicate that the effect of $a_{\rm k}$ and $\varepsilon$ in regulating the parameter space for multiple critical points appears to be analogous in nature. 

\section{Accretion solutions with shock transitions}

\label{sec:Global accretion solution contain shock}

\begin{figure}
	\includegraphics[width=\columnwidth]{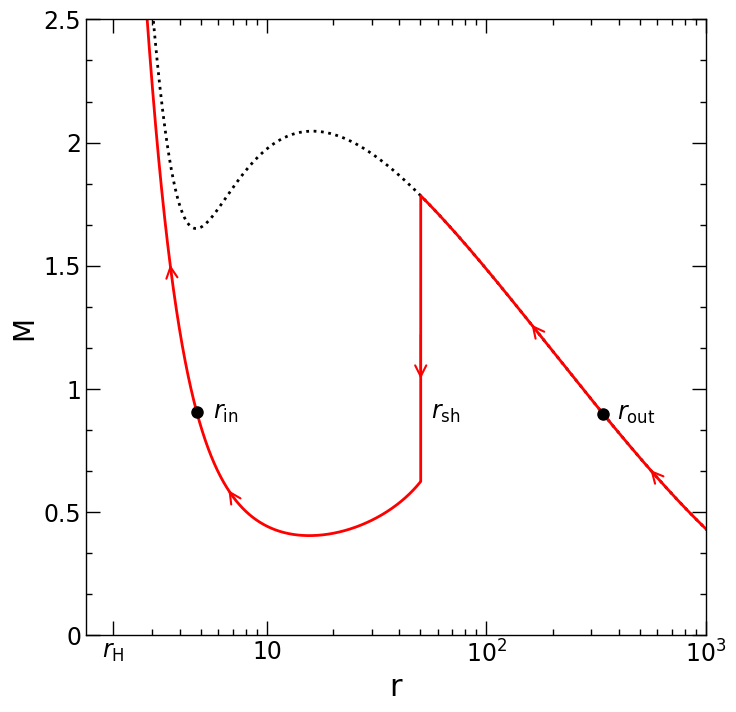}
	\caption{Example of a shock induced global accretion solution around black hole where the variation of Mach number ($M$) with the radial coordinate ($r$) is shown. The solution is obtained for $a_{\rm k} = 0$, $\varepsilon = 3$, $\lambda = 3.0$ and $E = 1.0005$. Vertical arrow indicates the location of the shock transition at $r_\text{sh} = 50.1706$. Arrows denote the overall flow motion towards the black hole. See text for details. 
	}
	\label{fig:shock_solution}
\end{figure}

In this section, we focus on only to those accretion solutions that may contain shocks. In an accretion process, flow starts accreting towards black hole subsonically from the outer edge of the disc. Depending on the input parameters, flow becomes supersonic after crossing the outer critical point ($r_{\rm out}$) and continues to accreate towards the horizon.  Meanwhile, flow experiences virtual barrier due to the centrifugal repulsion that eventually triggers the discontinuous transition of the flow variables, when relativistic shock conditions are satisfied \cite[]{Fukue_1987, Chakrabarti_1989, Yang_Kafatos_1995, Chakrabarti_Das_2004, Chattopadhyay_Kumar_2016, Kumar_Chattopadhyay_2017, Dihingia_et_al_2018, Dihingia_Das_Mandal_2018, Dihingia_Das_manda_2018b, Dihingia_Das_Nandi_2019, Dihingia_et_al_2019, Sen-2022}. Generally, the shocked-accretion solution is preferred over the shock-free solutions as the entropy associated to the former solution remain always higher at the inner part of the disk \cite{Becker_Kazanas_2001}. In order to calculate the location of the shock transition, we use the relativistic shock conditions which are given by \cite[]{Taub1948},
\begin{equation}
\label{eq:shock condition}
\begin{split}
&\left[ \rho u^{r} \right] = 0; \hspace{0.3cm} \left[ (e + p) u^{t}u^{r} \right] = 0; \\ &
\left[ (e + p) u^{r}u^{r} + pg^{rr} \right] = 0.
\end{split}
\end{equation}
In Eq.~(\ref{eq:shock condition}), the quantities within the square bracket refer the difference of their values across the shock front. 

In Fig. \ref{fig:shock_solution}, we present an example of a global accretion solutions comprising shock, where Mach number ($M$) is plotted with the radial distances ($r$). Here, we choose flow parameters as $(\lambda, E) = (3, 1.0005)$, and black hole parameters as $(a_{k},\varepsilon) = (0, 3)$. In the figure, the accretion solution is plotted using solid curve. In reality, accretion flow can smoothly enters into the black hole after crossing $r_{\rm out}=339.7504$ as indicated by the dotted (black) curve. However, flow finds a possibility of shock transition at $r_{\rm sh} = 50.1706$ and it jumps from supersonic to subsonic branch which is shown using vertical arrow. After the shock, flow again become supersonic after crossing $r_{\rm in} = 4.7916$ before falling into the black hole. Overall, arrows indicate the direction of flow motion towards the horizon from the outer edge of the disk.

\subsection{Shock properties}

\label{sec:shock_properties}

\begin{figure}
	\includegraphics[width=\columnwidth]{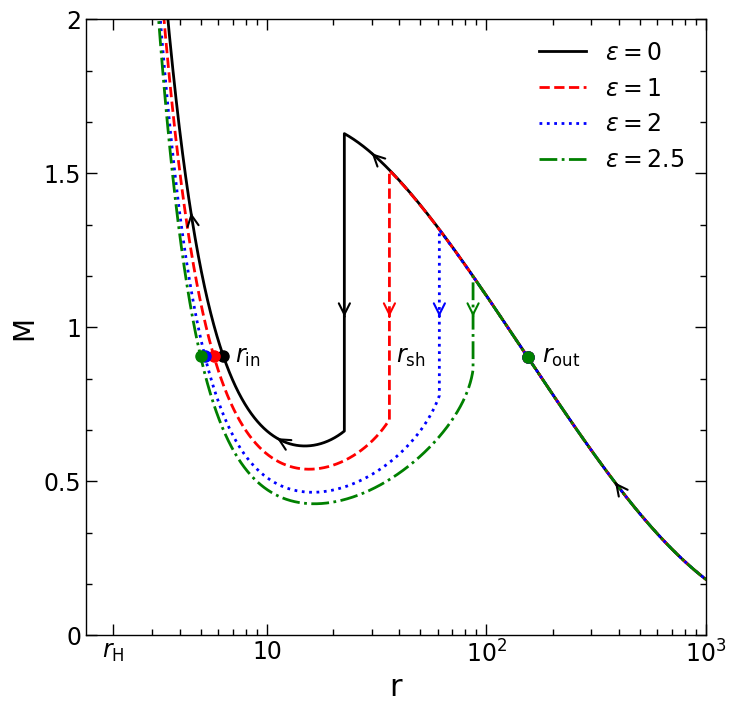}
	\caption{Variation of Mach number ($M$) with the radial coordinates ($r$) for different deformation parameters ($\varepsilon$). Here, we choose $a_{k} = 0$, $\lambda = 3$ and $E = 1.0013$. Vertical arrows indicate the radius of the shock transition at $r_\text{sh} =  22.5278, 36.1334, 61.0066$ and $86.8639$ corresponding to $\varepsilon =  0, 1, 2$ and $2.5$, respectively. Critical points ($r_\text{in}$ and $r_\text{out}$) are annotated by the filled circles. See text for details.
	}
	\label{fig:shock_solutions_different_epsilon}
\end{figure}

Fig. \ref{fig:shock_solutions_different_epsilon} shows the dynamical structure of the shocked accretion flow resulted due to the variation of spacetime deformation ($\varepsilon$). We consider accretion flow with energy $E = 1.0013$ and angular momentum $\lambda = 3.0$ that are injected from the outer edge of the disk at $r_{\rm edge} = 300$ towards the black hole of spin $a_{\rm k} = 0$. Here, we choose mass accretion rate $\dot{M} = 0.1 \dot{M}_{\rm Edd}$ with $M_{\rm BH} = 10M_{\odot}$, where $\dot{ M}_{\rm Edd} (= 1.39\times 10^{18}\frac{M_{\rm BH}}{M_{\odot}}$ gm s$^{-1}$) refers the Eddington accretion rate (the point is to note that we take energy conversion efficiency $\eta = 0.1$ in $\dot{M}_{\rm Edd}$) and $M_{\odot}$ denotes the solar mass. For $\varepsilon = 0$, flow encounters shock transition at $r_{\rm sh} = 22.5278$ and the solution is depicted using solid (black) curve. We now increase the spacetime deformation keeping other input parameters fixed, and observe that shock radius gradually recedes away from the black hole horizon. For $\varepsilon = 1, 2,$ and $2.5$, we obtain $r_{\rm sh} = 36.1334, 61.0066$ and $86.8639$, respectively and these solutions are shown using dashed (red), dotted (blue) and dot-dashed (green) curves. Needless to mention that indefinite increase of $\varepsilon$ is not possible because of the fact that beyond a critical limit, shock ceases to exist as the shock conditions (Eq.~\ref{eq:shock condition}) fail to satisfy.

\begin{figure}.
	\includegraphics[width=\columnwidth]{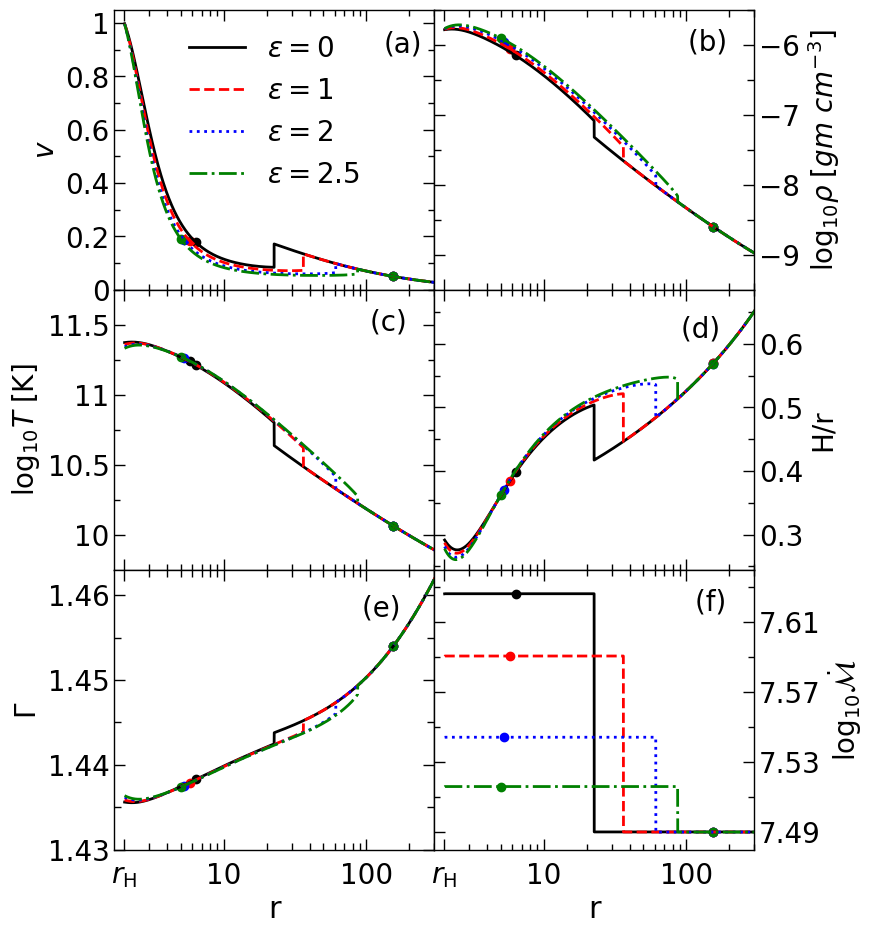}
	\caption{
		Variation of (a) radial velocity ($v$), (b) density ($\rho$), (c) temperature ($T$), (d) vertical scale-height ($H/r$), (e) adiabatic index ($\Gamma$) and (f) entropy accretion rate ($\dot{\mathcal{M}}$) as a function of radial coordinates ($r$) in a spacetime with different deformation parameters ($\varepsilon$). Each solid (black), dashed (red), dotted (blue) and dot-dashed (green) curves are used for $\varepsilon =  0, 1, 2$ and $2.5$, respectively. Here, we choose ($\lambda, E) = (3, 1.0013)$, and $a_{\rm k}=0$. In each panel, shock locations are indicated by the vertical lines at $r_{\rm sh} = 22.5278, 36.1334, 61.0066$ and $86.8639$, respectively. Critical points ($r_{\rm in}$ and $r_{\rm out}$) are marked by the filled circles. See text for details.
	}
	\label{fig:effect of epsilon on flow variables for shock solutions}
\end{figure}

In Fig. \ref{fig:effect of epsilon on flow variables for shock solutions}, we present the profile of various flow variables corresponding to the shock-induced global accretion solutions illustrated in Fig. \ref{fig:shock_solutions_different_epsilon}. In Fig. \ref{fig:effect of epsilon on flow variables for shock solutions}a, radial velocity ($v$) profile is plotted as function of radial coordinates ($r$) where discontinuous jump of $v$ is clearly seen. Solid (black), dashed (red), dotted (blue) and dot-dashed (green) curves denote the results obtained for $\varepsilon = 0, 1, 2$, and $2.5$, respectively. It is noteworthy that the difference of flow velocity across the shock front decreases with $\varepsilon$ and beyond a critical limit of $\varepsilon$, smooth transonic accretion solutions containing $r_{\rm out}$ only remains as shock disappears. We discuss the critical limit of $\varepsilon$ in the subsequent sections while studying the shock properties. In Fig. \ref{fig:effect of epsilon on flow variables for shock solutions}b, we show the variation of density ($\rho$) with $r$. Across the shock front, since the flow velocity drops down, the density of the post-shock flow (hereafter PSC) jumps to higher value. This happens simply due to preserve the conservation of mass flux across the shock front. Indeed, the density compression decreases as $\varepsilon$ is increased. We show the temperature ($T$) profile of the accretion flows in Fig. \ref{fig:effect of epsilon on flow variables for shock solutions}c, where we find that the temperature jumps suddenly at PSC. In reality, most of the kinetic energy of pre-shock flow is converted into the thermal energy after the shock transition and hence, the rise of post-shock temperature is observed as a consequence of PSC heating. We tabulate the temperature of the shocked accretion flows at different radii ($r_{\rm in}$, $r_{\rm out}$ and $r_{\rm sh}$) in Table \ref{tab:table-2}. In Fig. \ref{fig:effect of epsilon on flow variables for shock solutions}d, we present the radial dependence of the vertical scale-height ($H/r$). We find that accretion flow maintains $H/r < 1$ all throughout from the outer edge to the horizon even in the presence of shock transition. In Fig. \ref{fig:effect of epsilon on flow variables for shock solutions}e, we show the profile of adiabatic index ($\Gamma$) as function of $r$. As expected, $\Gamma$ anti-correlates with the flow temperature at all radii. Finally, in Fig. \ref{fig:effect of epsilon on flow variables for shock solutions}f, we depict the profile of entropy accretion rate ($\dot{\mathcal M}$) and observe that flow at PSC possesses high entropy content. With this, we wish to emphasize that the location of the shock eventually provides the size of the PSC that contains swarm of hot electrons. When soft photons from the pre-shock disk interact with these hot electrons via inverse Comptonization process, high energy radiations are produced which are commonly observed in Galactic black hole x-ray binaries (GBH-XRBs) \cite[and references therein]{Chakrabarti_Titarchuk1995,Nandi-etal2012,Nandi-etal2018}.

{\red 
\begin{table*}
	\centering
	\caption{Deformation parameter ($\varepsilon$), critical point locations ($r_{\rm in}, r_{\rm out}$), critical point temperatures ($T (r_{\rm in})$, $T (r_{\rm out})$), shock location ($r_{\rm sh}$), pre-shock temperatures ($T_{-} (r_{\rm sh}))$ and post-shock temperatures ($T_{+}(r_{\rm sh}))$ for shock-induced global accretion solutions presented in Fig. \ref{fig:effect of epsilon on flow variables for shock solutions}.}
	\label{tab:table-2}
	\begin{ruledtabular}
		\begin{tabular}{lccccccc}
			$\varepsilon$ & $r_{\rm in}$ & $r_{\rm out}$ & $T (r_{\rm in})$ & $T (r_{\rm out})$ & $r_{\rm sh}$ & $T_{-} (r_{\rm sh})$ & $T_{+} (r_{\rm sh})$\\
			 & & & $(\times ~ 10^{10}\rm K)$ & $(\times ~ 10^{10}\rm K)$ &  & $(\times ~ 10^{10}\rm K)$ & $(\times ~ 10^{10}\rm K)$ \\
			\hline
			 0 & 6.3388 & 154.2915 & 16.2554 & 1.1567 & 22.5278 & 4.3280 & 6.3134\\
			 1 & 5.7364 & 154.2562 & 17.3149 & 1.1568 & 36.1334 & 3.0595 & 4.1752\\
			 2 & 5.2186 & 154.2115 & 18.2422 & 1.1571 & 61.0066 &  2.1224 & 2.5985\\
			 2.5 & 4.9869 & 154.1944 & 18.6363 & 1.1572 & 86.8639 & 1.6735 & 1.8726\\
		\end{tabular}
	\end{ruledtabular}
\end{table*}
}

\begin{figure}
	\includegraphics[width=\columnwidth]{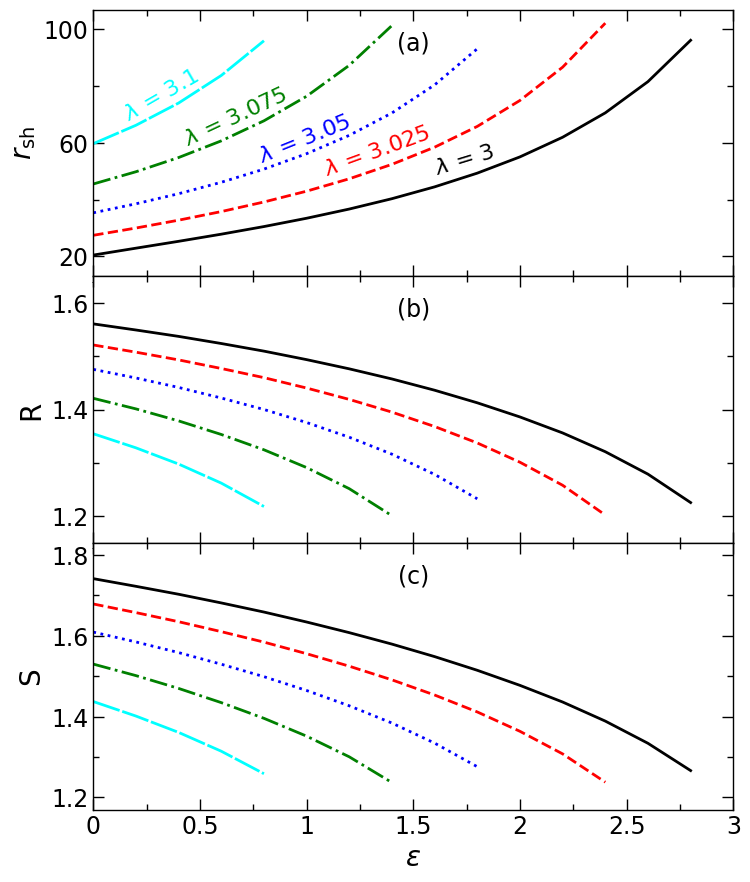}
	\caption{Variation of (a) shock location ($r_\text{sh}$), (b) compression ration ($R$) and (c) shock strength ($S$) with the deformation parameters ($\varepsilon$). Solid (black), dashed (red), dotted (blue), dot-dashed (green) and big-dashed (cyan) curves denote results for $\lambda = 3, 3.025, 3.05, 3.075$ and $3.1$, respectively. Here, we choose $a_{\rm k}=0$ and $E = 1.0012$. See text for details.}
	\label{fig:shock_properties}
\end{figure}

In the next, we study the properties of shocks in terms of the input parameters. For that we consider accretion flow with $E = 1.0012$ and calculate shock radius ($r_{\rm sh}$) as function of $\varepsilon$ for a set of angular momentum ($\lambda$). Here, we choose $a_{\rm k}=0$. The obtain results are presented in Fig. \ref{fig:shock_properties}a, where solid (black), dashed (red), dotted (blue), dot-dashed (green), and big-dashed (cyan) curves are for $\lambda = 3.0, 3.025, 3.05, 3.075$, and $3.1$, respectively. We find that for a fixed $\lambda$, $r_{\rm sh}$ increases with $\varepsilon$, however, shock is seen to disappear when $\varepsilon$ exceeds its critical value $\varepsilon^{\rm cri}$. It is evident from the figure that for a fixed $E$, $\varepsilon^{\rm cri}$ decreases as $\lambda$ is increased. Therefore, for relatively lower $\lambda$, the possibility of obtaining the shocked-induced global accretion solutions is very much likely even the strength of deformation is higher and vice versa. We also notice that for given $\varepsilon$, $r_{\rm sh}$ settles down at larger radius for flows with higher $\lambda$. This clearly suggests that shocks under consideration are centrifugally drive. Since both density and temperature of the accreting flow are increased substantially at PSC, it is therefore worthy to study the compression ratio ($R$) and shock strength ($S$). In reality, compression ratio ($R=\Sigma_{+}/\Sigma_{-}$, `$-$' and `$+$' refer quantities just before and after the shock transition) measures the density compression across the shock front and in Fig. \ref{fig:shock_properties}b, we show the variation of $R$ as function of $\varepsilon$ for the same set of input parameters as in Fig. \ref{fig:shock_properties}a. As expected, $R$ decreases with the increase of $\varepsilon$. This happens because the density compression at PSC generally weakens as shock recedes outward (see Fig. \ref{fig:effect of epsilon on flow variables for shock solutions}b). In Fig.~\ref{fig:shock_properties}c, we depict the variation of $S$ (defined as $S=M_{-}/M_{+}$  and is a measure of temperature jump at PSC) with $\varepsilon$ corresponding to the results presented in Fig. \ref{fig:shock_properties}a. We observe that $S$ decreases with $\varepsilon$ for a given $\lambda$ which agrees with our previous findings (see Fig. \ref{fig:effect of epsilon on flow variables for shock solutions}c).

\begin{figure}
	\includegraphics[width=\columnwidth]{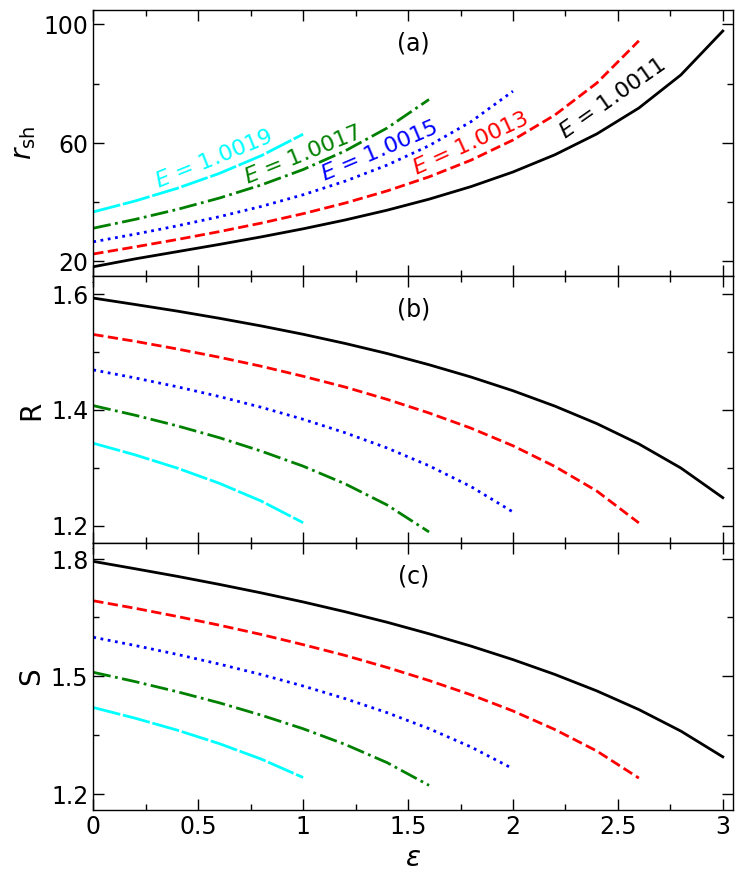}
	\caption{Variation of (a) shock location $r_\text{sh}$, (b) compression ration ($R$) and (c) shock strength ($S$) as a function of the deformation parameters ($\varepsilon$). Solid (black), dashed (red), dotted (blue), dot-dashed (green) and big-dashed (cyan) curves are for $E = 1.0011, 1.0013, 1.0015, 1.0017$ and $1.0019$, respectively. Here, we fix $a_{\rm k} = 0$ and $\lambda = 3$. See text for details.}
	\label{fig:shock_properties1}
\end{figure} 

In Fig.~\ref{fig:shock_properties1}, we examine the shock properties for flows having $\lambda = 3.0$ but different energies $E$. Here, we set $a_{k} = 0$. In panels (a-c), we present the variation of $r_{\rm sh}$, $R$ and $S$ with $\varepsilon$, where solid (black), dashed (red), dotted (blue), dot-dashed (green), and big-dashed (cyan) curves are for $E = 1.0011, 1.0013, 1.0015, 1.0017$, and $1.0019$, respectively. We find that for a fixed $E$, $r_{\rm sh}$ increases with $\varepsilon$ and beyond $\varepsilon<\varepsilon^{\rm cri}$ shock disappears. We notice that for a given $\lambda$, $\varepsilon^{\rm cri}$ decreases for flows with higher energies. It is worth mentioning that $\varepsilon^{\rm cri}$ does not bear universal values as it explicitly depends on the other input parameters. We also find that for a fixed $\varepsilon$, flow experiences shock transition ($r_{\rm sh}$) at larger radii when energy ($E$) is increased. Further, we observe that for a given set of ($\lambda, E$), both $R$ and $S$ decreases with the increase of $\varepsilon$. This happens because the enhanced $\varepsilon$ pushes the shock front outwards that reduces the overall density compression as well as temperature jump at PSC.

\subsection{Parameter space for shocks}

\label{sec:Shock parameter space}

\begin{figure}
	\includegraphics[width=\columnwidth]{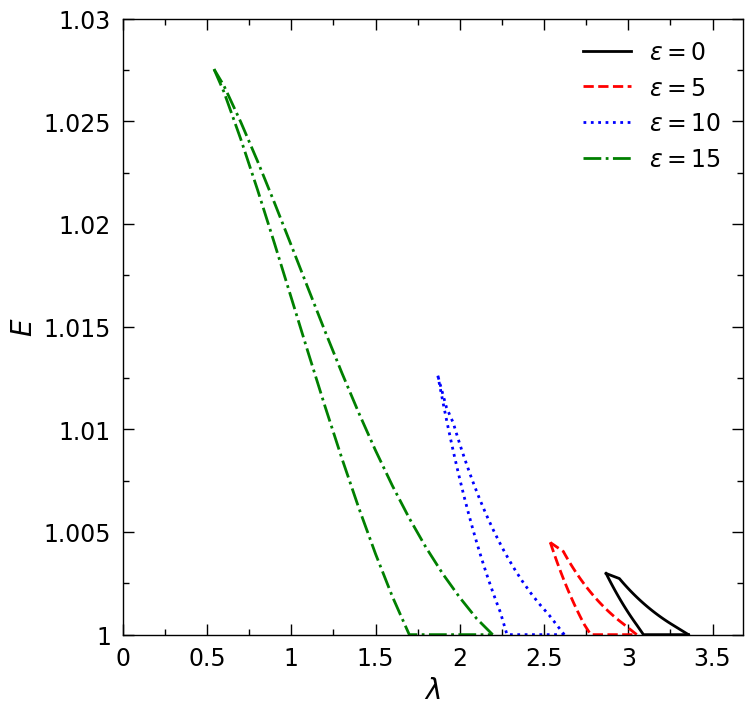}
	\caption{Modification of the shock parameter space in $\lambda-E$ plane as a function of deformation parameter ($\varepsilon$). Here, we fix $a_{\rm k}=0.0$. Regions bounded with solid (black), dashed (red), dotted (blue) and dot-dashed (green) curves are obtained for $\varepsilon = 0, 5, 10$ and $15$, respectively. See text for details.}
	\label{fig:parameter_space_shock_1}
\end{figure}

We already indicate that shock-induced global accretion solutions are not isolated solutions, in fact they exist for wide range of input parameters. In order to quantify the allowed range of input parameters, we separate the effective region of the parameter space in $\lambda -E$ plane that admits shocked accretion solutions. Towards this, in Fig. \ref{fig:parameter_space_shock_1}, we show the modification of the shock parameter space due to the increase of $\varepsilon$, where regions bounded by the solid (black), dashed (red), dotted (blue) and dot-dashed (green) are obtained for $\varepsilon = 0, 5, 10$ and $15$, respectively. Here, we choose $a_{\rm k} = 0$. The solid curve depicted the region that agrees with Fig.~5 from \citet[]{Dihingia_Das_Nandi_2019}. We note that the domain of the parameter space for shock gradually increases with the increase of $\varepsilon$ and shifts towards the higher energy and lower angular momentum sides. This findings clearly indicate that the possibility of shock formation is eventually increased as the spacetime deformation is increased. In addition, low angular momentum flow around black hole seems to possess standing shock provided the level of spacetime deformation is relatively high and vice versa.

\begin{figure}
	\includegraphics[width=\columnwidth]{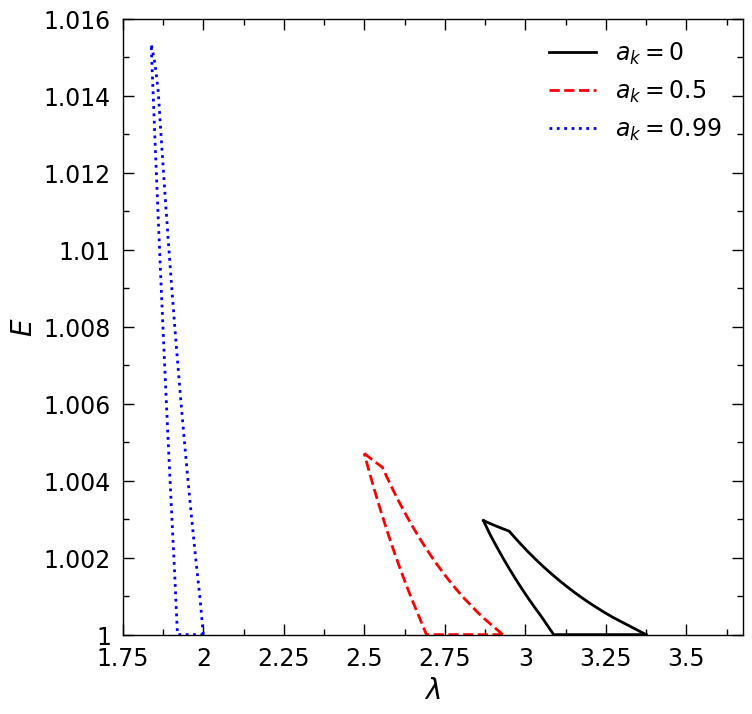}
	\caption{Modification of shock parameter space in $\lambda-E$ for different Kerr parameters ($a_{\rm k}$). Here, we choose $\varepsilon = 0.02$. Effective regions bounded using solid (black), dashed (red) and dotted (blue) curves are obtained for $a_{\rm k} = 0, 0.5$ and $0.99$, respectively. See text for details.}
	\label{fig:parameter_space_shock_3}
\end{figure}

Next, we intend to examine the role of black hole spin ($a_{\rm k}$) in modifying the effective region of parameter space in ($\lambda-E$) plane for
standing shock. In Fig. \ref{fig:parameter_space_shock_3}, we show the results where shock parameter space is calculated for $\varepsilon = 0.02$ considering different $a_{\rm k}$. The effective regions bounded with solid (black), dashed (red) and dotted (blue) curves correspond to $a_{\rm k}=0, 0.5$, and $0.99$, respectively. We find that the allowed region for the standing shock solution shifts toward lower angular momentum as $a_{\rm k}$ increases. This shift occurs because the marginally stable angular momenta of the accreting material goes down when $a_{\rm k}$ is increased \cite[]{Das-Chakrabarti2008}.

\subsection{Zero angular momentum flow}

\label{sec:flow solutions associated with l = 0}

\begin{figure}
	\includegraphics[width=\linewidth]{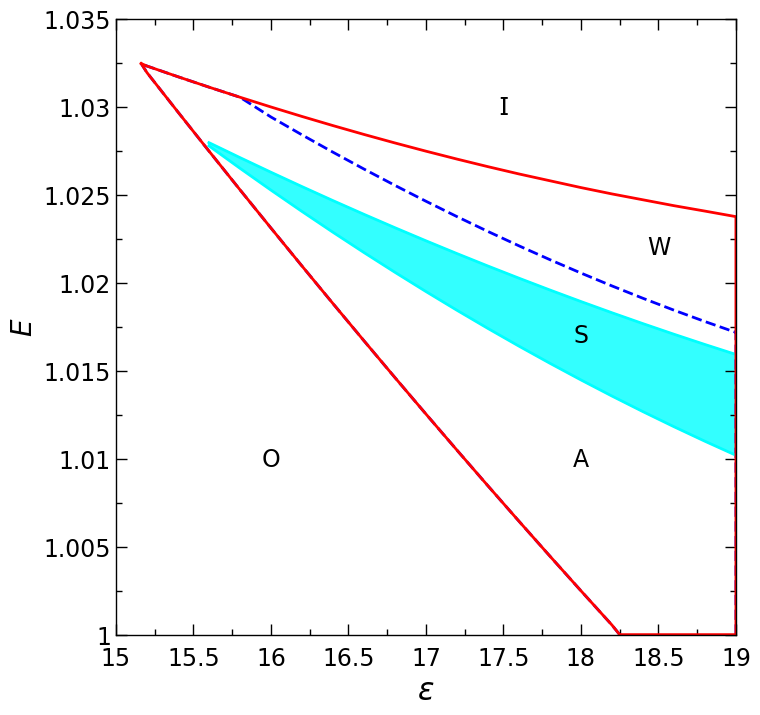}
	\caption{Separation of parameter space in the $\varepsilon-E$ plane according to the nature of flow solutions (O, A, W, S, I). Here, we choose $\lambda = 0.0$ and $a_{\rm k}=0.0$. See text for details.}
	\label{fig:parameter space for l = 0}
\end{figure}

In this section, we study the zero angular momentum flow (ZAMF, $\lambda=0$) in deformed Kerr spacetime. While doing this, we separate the region in the $\varepsilon-E$ plane according to the nature of the accretion solutions (see Fig. \ref{fig:division_parameter_space}). In Fig. \ref{fig:parameter space for l = 0}, we depict the obtained results for $\lambda = 0$ and $a_{k} = 0$, where the region bounded using red curves are obtained for multiple critical points. This region is further sub-divided into two domains using dashed (blue) curves ($\mathcal{\dot{M}_{\rm in}} = \mathcal{\dot{M}_{\rm out}}$), namely A and W, respectively. The shaded region (cyan) marked as S admits shock-induced global accretion solutions for ZAMF. The remaining regions marked as A and I allow accretion solutions that possess single critical points ($r_{\rm in}$ for I, and $r_{\rm out}$ for O). 

\begin{figure*}
	\includegraphics[width=0.8\linewidth]{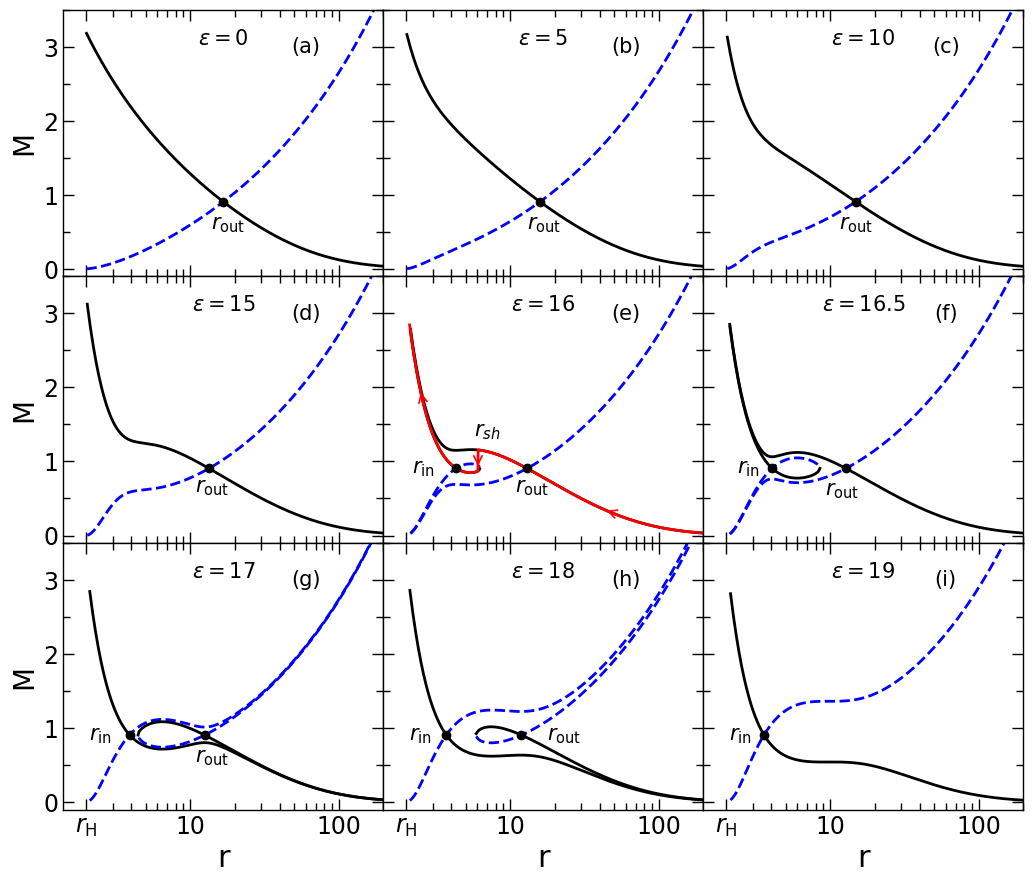}
	\caption{Variation of Mach number ($M$) with the radial distances ($r$) for zero angular momentum ($\lambda=0$) flows. Here, we set $E=1.0255$ and $a_{k}=0$. Results depicted in panels (a) to (i) correspond to smooth variation of $\varepsilon$ marked in respective panels. Solid (black) and dashed (blue) curves denote the accretion and wind solutions, respectively. Critical points ($r_{\rm in}, r_{\rm out}$) are marked using filled circles. Arrows (red) indicate the direction of flow motion for shock-induced global accretion solution (solid, red line) is shown where we obtain $r_\text{sh} = 6.0645$. See text for details.}
	\label{fig:flow solutions associated with l = 0 and E = 1.0255}
\end{figure*} 

Next, we examine how the accretion solutions are modified due to the change of $\varepsilon$ for ZAMF. In Fig. \ref{fig:flow solutions associated with l = 0 and E = 1.0255}, we depict the obtained results where Mach number ($M$) is plotted as function of radial coordinate ($r$). Here, we choose $\lambda = 0$, $E=1.0255$ and $a_{\rm k}= 0$. In each panel, solid (black) curve denotes accretion solution whereas dashed (blue) curve refers the corresponding wind branch.  In panel (a), (b), (c) and (d), we consider $\varepsilon = 0, 5, 10$ and $15$ that yield only outer critical points at $r_{\rm out}=16.7040, 15.8656, 14.8279$ and $13.3763$, respectively. As the spacetime deformation is increased further to $\varepsilon = 16$, we observe that the inner critical point appears at $r_{\rm in} = 4.3305$ along with $r_{\rm out} = 12.9792$, which is shown in panel (e). We further find $\dot{\cal M}_{\rm in} > \dot{\cal M}_{\rm out}$ and observe that standing shock conditions are satisfied at $r_{\rm sh} = 6.0645$ as shown using vertical arrow. The compression ratio and shock strength for this shock jump are calculated as $R=1.2408$ and $S=1.2677$, respectively. Needless to mention that the formation of shock in ZAMF is evidently resulted due to the spacetime deformation under consideration. The arrows indicate the overall direction of the flow motion. Here, we wish to emphasize that for the first time to our knowledge, we obtain the zero angular momentum shocked accretion solution around black holes embedded in deformed Kerr spacetime. In panel (f), flow solutions are plotted for $\varepsilon = 16.5$, where the critical points are calculated as $(r_{\rm in}, r_{\rm out}) = (4.0617, 12.7550)$. Although flow contains multiple critical points and $\dot{\mathcal M}_{\rm in} > \dot{\mathcal M}_{\rm out}$, shock transition does not happen as shock conditions are not satisfied. For $\varepsilon = 17$, accretion flow still possesses multiple critical points, however, changes its character as shown in Fig. \ref{fig:flow solutions associated with l = 0 and E = 1.0255}g. We notice that the solution passing through $r_{\rm out} = 12.5073$ no longer connects the horizon ($r_{\rm H}$) with the outer edge of the disk ($r_{\rm edge}$), instead the solution passing through $r_{\rm in} = 3.9067$ smoothly connects $r_{\rm H}$ and $r_{\rm edge}$. Moreover, we get $\dot{\cal M}_{\rm in} < \dot{\cal M}_{\rm out}$ that disfavors the shock transition. For further increase of $\varepsilon = 18$, we obtain qualitatively similar solution as shown in panel (h), but the critical points are shifted as $(r_{\rm in}, r_{\rm out}) = (3.7102, 11.9028)$. For the limiting value of $\varepsilon = 19$, outer critical point disappears leaving only the inner critical point at $r_{\rm in} = 3.5821$ as depicted in Fig. \ref{fig:flow solutions associated with l = 0 and E = 1.0255}i. Solutions of this kind resemble the advection-dominated accretion flows (ADAF) as reported in earlier works \cite[]{Chakrabarti_1989,Chakrabarti_1990,narayan_et_al_1997}. With this, we indicate that spacetime deformation plays viable role in determining the overall nature of the accretion solutions around black hole including shocks. We summarize the critical point locations and the nature of the accretion solutions by means of $\varepsilon$ in Table~\ref{tab:table-3}.  	

\begin{table*}
	\centering
	\caption{Deformation parameters ($\varepsilon$), inner critical points ($r_{\rm in}$) and orbit, outer critical points ($r_{\rm out}$) and orbit, shock location ($r_{\rm sh}$), flow types for accretion solutions presented in Fig. \ref{fig:flow solutions associated with l = 0 and E = 1.0255}.}
	\label{tab:table-3}
	\begin{ruledtabular}
		\begin{tabular}{lcccccc}
			$\varepsilon$ & $r_{\rm in}$~(saddle) & Orbits & $r_{\rm out}$~(saddle) & Orbits & $r_{\rm sh}$ & Types\\
			\hline
			0 & --- & --- & $16.7040$ & Heteroclinic & --- & O\\
			 5 & --- & --- & $15.8656$ & Heteroclinic & --- & O\\
			 10 & --- & --- & $14.8279$ & Heteroclinic & --- & O\\
			 15 & --- & --- & $13.3763$ & Heteroclinic & --- & O\\
			16 & $4.3305$ & Homoclinic & $12.9792$& Heteroclinic & $6.0645$ & A, S\\
			 16.5 & $4.0617$ & Homoclinic & $12.7550$ &  Heteroclinic & --- & A\\
			 17 & $3.9067$ & Heteroclinic & $12.5073$ & Homoclinic  & --- & W\\
			 18 & $3.7102$ & Heteroclinic & $11.9028$ & Homoclinic & --- & W\\
			 19 & $3.5821$ & Heteroclinic & --- & ---& --- & I\\
		\end{tabular}
	\end{ruledtabular}
\end{table*}

\section{Naked Singularity in deformed Kerr spacetime}

\label{sec:naked_singularity} 

In deformed Kerr spacetime, the flow solutions around central object behave differently as the suitable combination of $\varepsilon$ and $a_{k}$ yields naked singularity. Towards this, we employ the horizon condition as $g^{\rm rr}=g_{\rm rr}^{-1}=0$ and upon simplification, we get
\begin{equation}
\label{eq:event horizon}
r_{\rm H}^{5} - 2r_{\rm H}^{4} + a_{k}^{2}r_{\rm H}^{3} + a_{k}^{2}\varepsilon = 0,
\end{equation}
where $r_{\rm H}$ denotes event horizon. For a suitable choice of ($a_{\rm k}, \varepsilon$), when Eq.~(\ref{eq:event horizon}) does not provide any real roots, the central object evidently represents the naked singular spacetime, instead of black hole. Considering such scenario, we study the properties of the hydrodynamic flow around the naked singularity and examine how $\varepsilon$ and $a_{k}$ regulate the nature of the accretion solutions. While doing so, we follow the same methodologies as discussed in Section~\ref{sec:Hydrodynamics with deformation}.

\subsection{Critical point properties}

\label{sec:critical points analysis for naked singularity}

\begin{figure}
	\includegraphics[width=\columnwidth]{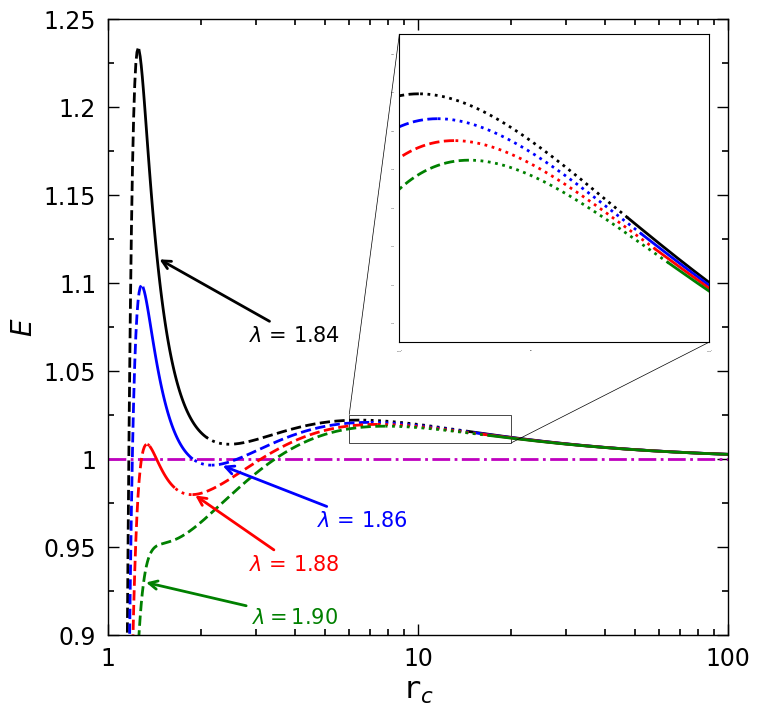}
	\caption{Variation of flow energy ($E$) with the critical points ($r_{c}$) for different angular momentums ($\lambda$). Here, we choose $a_{\rm k} = 0.99$ and $\varepsilon = 0.03$. Saddle, nodal and O-type critical points are indicated with the solid, dotted and dashed curves. Horizontal line (dot-dashed) is plotted at specific energy $E=1$. A part of the plot is zoomed for the purpose of clarity. See text for details.}
	\label{fig:E vs. r for naked singularity}
\end{figure}

In Fig. \ref{fig:E vs. r for naked singularity}, we present the variation of the flow energy ($E$) with $r_c$ for $(a_{\rm k}, \varepsilon) = (0.99,0.03)$ that renders naked singularity. The obtained results plotted using black, blue, red and green curves are for $\lambda = 1.84, 1.86, 1.88$ and $1.90$, respectively. For each $\lambda$, solid, dotted and dashed curves represent  saddle, nodal and spiral (or O-type) critical points. We observe that unlike black hole case, there exists four critical points in an energy range ($E\ge1$) that eventually depends on the angular momentum of the flow ($\lambda$). On the contrary, when $E<1$, flow may have single or at most three critical points depending on $\lambda$. Interestingly, we note that the critical points located closest to the central singularity are of spiral type and hence they are not physical as accretion flow can not pass through them. For a given $\lambda$, the critical points are in general arise in sequence starting from the inner most one as spiral-saddle-nodal-spiral-nodal-saddle, provided flow possesses multiple critical points.

\subsection{Flow solutions of different kinds}

\label{sec:global accretion solutions naked singularity}

\begin{figure*}
	\includegraphics[width=\linewidth]{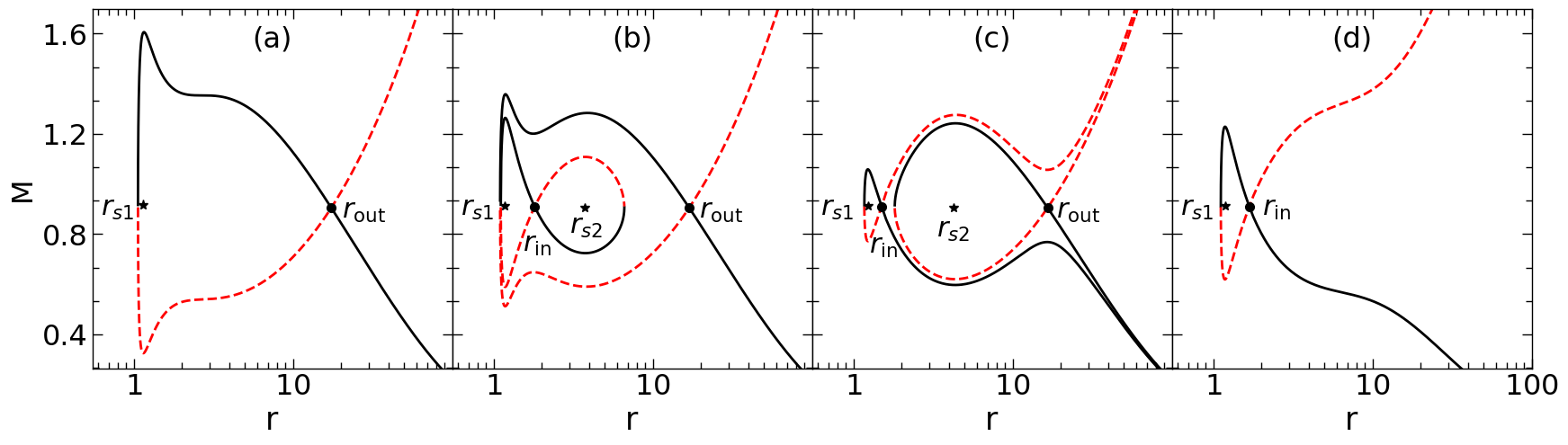}
	\caption{Variation of Mach number ($M = v/C_{s}$) with the radial distances ($r$) for $a_{\rm k}=0.99$ and $\varepsilon=0.03$. In each panel, solid (black) and dashed (red) curves represent the accretion and wind solutions, respectively. Filled circles and asterisks denote saddle and O-type critical points, respectively. Here, we choose $(\lambda, E) = (1.82, 1.0137)$, $(1.85,1.0137)$, $(1.87,1.0137)$, and $(1.85,1.025)$ in respective panels (a-d). See text for details.}
	\label{fig:global accretion solutions naked singularity}
\end{figure*}

In Fig. \ref{fig:global accretion solutions naked singularity}, we present all possible type of flow solutions around naked singularity, where the variation of Mach number ($M$) as function of radial distance ($r$) is depicted in each panel. Here, we choose $a_{\rm k} = 0.99$ and $\varepsilon = 0.03$ that yields naked singularity. In panel (a), we fix $\lambda = 1.82$ and $E = 1.0137$, and obtain two critical points; the closest one from the singularity is of spiral type formed at $r_{s1}=1.1493$ (filled asterisk) and the furthest one is of saddle type located at $r_{\rm out}=17.4339$ (filled circle). We calculate the flow solutions passing through $r_{\rm out}$ and plot the accretion and wind branches using solid and dashed curves, respectively. We note that during the course of accretion, rotating transonic flow usually piles up and hence tends to co-rotate along a surface close to the naked singularity usually known as {\it naked surface} \cite[][]{Dihingia_et_al_2020}. This feature is clearly seen in panel (a) as $M$ for accretion drops down around the singular point. In Fig.~\ref{fig:global accretion solutions naked singularity}b, we show the flow solutions for $\lambda = 1.85$ and $E = 1.0137$, where four critical points are obtained. Among them, two are saddle type ($r_{\rm in} = 1.8155$ and $r_{\rm out} = 16.9517$ marked as filled circles) and remaining two are spiral type ($r_{s1} = 1.1832$ and $r_{\rm s2} = 3.7319$ marked as filled asterisks). As before, solid and dashed curves denote the accretion and wind branches. We notice that solution passing through $r_{\rm out}$ connects the {\it naked surface} and the outer edge of the disk ($r_{\rm edge}$), however, flow possesses $r_{\rm in}$ does not extend up to $r_{\rm edge}$, instead it becomes closed in between $r_{\rm in}$ and $r_{\rm out}$. Moreover, we find $\dot{\mathcal M}_{\rm in} > \dot{\mathcal M}_{\rm out}$. In panel (c), we choose $\lambda = 1.87$ and $E= 1.0137$ and again obtain two saddle and two spiral critical points at $r_{\rm in} = 1.5073$, $r_{\rm out} = 16.6046$, $r_{s1}=1.2339$, and $r_{s2} = 4.2651$, respectively. Here we observe that the overall nature of the flow solutions is changed compared to the solutions presented in (b) and $\dot{\mathcal M}_{\rm in} < \dot{\mathcal M}_{\rm out}$. In fact, solution passing through $r_{\rm out}$ becomes closed, but the same containing $r_{\rm in}$ smoothly connects {\it naked surface} and $r_{\rm edge}$. In panel (d), we set $\lambda =1.85$ and $E= 1.025$, and find $r_{s1}=1.1856$ (filled asterisk) and $r_{\rm in} = 1.6893$ (filled circle). As before, solid and dashed curves passing through $r_{\rm in}$ denote the accretion and wind branches. Finally, we wish to mention that all these flow solutions are in agreement with the results reported in  \cite{Dihingia_et_al_2020}.

\subsection{Deformation parameter ($\varepsilon$) separating BH and NS}

\label{sec:a vs epsilon} 

\begin{figure}
	\includegraphics[width=\columnwidth]{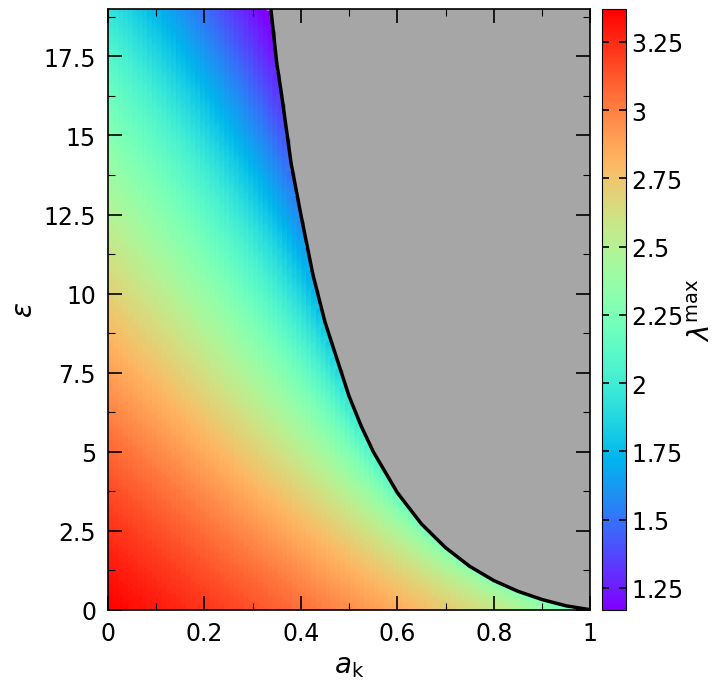}
	\caption{Parameter space in $a_{\rm k}-\varepsilon$ plane that admits flow solutions. Thick solid curve separates the effective domains for BH and NS, respectively. In BH region, color map denotes the 2D projection of 3D plot of $a_{\rm k}, \varepsilon$ and maximum angular momentum $\lambda^\text{max}$. The vertical color bar indicates the range of $\lambda^\text{max}$. See text for details.}
	\label{fig:a vs epsilon}
\end{figure}

In this section, we intend to determine the range of deformation parameter ($\varepsilon$) that yields the black hole spacetime. While doing this, for a given $a_{\rm k}$, we freely vary $\varepsilon$, $\lambda$, and $E~(>1)$ that provides a maximum limit of $\varepsilon~(= \varepsilon^{\rm max})$ such that the flow from the outer edge of the disk smoothly accretes onto the black hole (see inset panels of Fig. \ref{fig:division_parameter_space}). When $\varepsilon > \varepsilon^{\rm max}$, the accreting matter starts to pile up around the O-type critical points very close to the central singularity (see Fig. \ref{fig:global accretion solutions naked singularity}) resulting the inaccessible region (called naked surface) to the flows. With this, in Fig. \ref{fig:a vs epsilon}, we present the obtained results where solid (black) curve denotes the upper limit of deformation parameter ($\varepsilon^{\rm max}$) as function of $a_{\rm k}$ that separates the domain of black holes (shaded in rainbow color) from the naked singularities (shaded in gray). This findings are in agreement with $\varepsilon^{\rm max}(a_{\rm k})$ which is obtained by solving Eq.~(\ref{eq:event horizon}). In the figure, color code denotes the maximum angular momentum ($\lambda^{\max}$)  corresponding to a given set of ($a_{\rm k}, \varepsilon$) that admits closed solution passing through $r_{\rm in}$ (see panel (A) of Fig. \ref{fig:division_parameter_space}). The colorbar indicates the range of $\lambda^\text{max}$. We find that for $\varepsilon \rightarrow 0$, $\lambda^{\max}$ gradually decrease with the increase of $a_{\rm k}$ which is consistent with Figure 5 of \cite{Dihingia_Das_Nandi_2019}. On the contrary, we note that for a given $a_{\rm k}$, $\lambda^{\max}$ decreases with the increase of $\varepsilon$. Here, we restrict our analysis to the observational limit of the deformation parameter $\varepsilon \le 19$ \cite[]{Atamurotov_2013} and observe that the present analysis carried out based on accretion theory appears is consistent with \cite{Johannsen_Psaltis_2011,Bambhaniya_et_al_2021}. With this, we emphasize that accretion phenomenon offers an alternative approach in distinguishing the nature of the central source embedded in deformed Kerr spacetime. Moreover, since the accretion solutions successfully delineate the observational findings of extremely gravitating objects, we also infer that the present formalism would be immensely useful in explaining the astrophysical sources.

\section{Summary and conclusions}

\label{sec:Conclusions}

In this paper, we study the structure of a relativistic, inviscid, accretion flow in the JP spacetime \cite[]{Johannsen_Psaltis_2011} that describes the compact gravitating object embedded in deformed spacetime. We solve the conservation equations that governs the dynamics of the accretion flow around the central object and examine the role of spacetime deformation ($\varepsilon$) in controlling the global accretion solutions in presence and absence of shock waves. We note that the spacetime geometry under consideration represents either BH or NS depending on the spacetime parameters ($a_{k}, \varepsilon$). We summarize our findings below.

We find that depending on the input parameters, namely energy ($E$), angular momentum ($\lambda$), spin parameter ($a_{\rm k}$), and deformation parameter ($\varepsilon$), flow may contain either single or multiple critical points around BH and NS (see Fig. \ref{fig:E_vs_R_different_l_and_epsilon} and Fig. \ref{fig:E vs. r for naked singularity}). We obtain the flow solutions containing single critical point and find that for increasing $\varepsilon$, the nature of the solutions changes as it possesses multiple critical points (see Fig. \ref{fig:M_vs_r_different_epsilon} and Fig. \ref{fig:global accretion solutions naked singularity}). We identify the effective domain of the parameter space in $\lambda-E$ plane for multiple critical points which is further sub-divided based on the entropy criteria $i.e.$, $\dot{\mathcal M}_{\rm in} \lessgtr \dot{\mathcal M}_{\rm out}$, $r_{\rm in}$ and $r_{\rm out}$ being the inner and outer critical points, respectively (see Fig. \ref{fig:division_parameter_space}). Further, we classify the multiple critical point parameter space in terms of both $\varepsilon$ and $a_{\rm k}$, and find that parameter space strongly depends on them (see Fig. \ref{fig:comparison_parameter_space_fixed_a}  and Fig. \ref{fig:comparison_parameter_space_fixed_epsilon}). Accretion flows containing multiple critical points are of special interest as they may contain shock waves and shock-induced global accretion solution is perhaps essential to understand the observational signatures of black holes \cite[]{Chakrabarti-Manickam2000,nandi2001source,Nandi-etal2012,Radhika-Nandi2014,Iyer-etal2015,Das-etal2021}.

One of the aims of the present paper is to calculate the global shocked accretion solution in deformed spacetime and examine the shock dynamics as a function of $\varepsilon$. We find that for flows with fixed input parameters, shock settles down at larger radius as the $\varepsilon$ is increased (see Fig. \ref{fig:shock_solutions_different_epsilon}). Since the shock location provides the size of PSC and PSC inverse Comptonizes the soft photons from the pre-shock flow to produce high energy radiations, it is therefore worthy to examine the shock properties as they are likely to decide the nature of emitted photons from PSC. Accordingly, we examine the variation of shock location ($r_{\rm sh}$), compression ratio ($R$) and shock strength ($S$) as function of input parameters (see Fig. \ref{fig:shock_properties} and Fig. \ref{fig:shock_properties1}).

We separate the region of the parameter space in $\lambda-E$ plane that admits standing shock. We find that for $a_{\rm k}=0$, as $\varepsilon$ is increased, the effective region of the parameter space is increased and shifted to lower angular momentum and higher energy sides. This suggests that the possibility of shock formation is enhanced for higher $\varepsilon$ provided relativistic shock conditions are satisfied (see Fig. \ref{fig:parameter_space_shock_1}). Similarly, for fixed $\varepsilon$, we obtain shock parameter space at relative higher angular momentum when $a_{\rm k}$ is small and vice versa (see Fig. \ref{fig:parameter_space_shock_3}). 

Further, for the first time to our knowledge, we report that zero angular momentum flow (ZAMF, $\lambda=0$) can harbor standing shock around BHs embedded in deformed Kerr spacetime. We find that such shocked accretion solutions are possible when the spacetime deformation is sufficiently high (see Fig. \ref{fig:parameter space for l = 0}). Accordingly, we infer that the spacetime deformation plays the pivotal role for the formation of standing shock in ZAMF (see Fig. \ref{fig:flow solutions associated with l = 0 and E = 1.0255}).

We also observe that the nature of the central source in deformed Kerr spacetime may yield as naked singularity (NS). This usually happens when the horizon condition (equation \ref{eq:event horizon}) does not provide real roots for a given set of ($a_{\rm k}, \varepsilon$). Considering this, we calculate the flow solutions involving either single and/or multiple critical points (see Fig. \ref{fig:global accretion solutions naked singularity}). Moreover, analyzing the accretion dynamics, we separate the domain of BH from NS in $a_{\rm k}-\varepsilon$ plane and find that obtained results are in agreement with \citet[]{Johannsen_Psaltis_2011,Bambhaniya_et_al_2021}. With this, we infer that the present formalism offers an alternative approach to examine the nature of the central source (either BH or NS) embedded in deformed Kerr spacetime (see Fig. \ref{fig:a vs epsilon}).

Few comments may be worthy to mention. Although a similar observation was reported in the case of Kerr-Taub-NUT (KTN) metric \cite{Dihingia_et_al_2020, Sen-2022}, the accretion solutions obtained from KTN metric and JP metric can not be compared on the same footing. The fact is that KTN metric is a solution of a particular gravitation equation, whereas the exact gravitation dynamics that leads to JP metric remains unclear till date. In reality, the JP metric is regarded as the deformation (parameterized with $\varepsilon$) of the usual Kerr solution. Moreover, the nut parameter ($n$) in KTN metric is a hair of the black hole and hence it has an extra macroscopic parameter on top of source mass and spin. On the contrary, JP metric does not contain any such extra hair except mass and spin. Accordingly, $\varepsilon$ is not introduced as a new macroscopic entity of the black hole, instead it quantifies the deformation of the usual Kerr spacetime. Under these circumstances, the nut parameter ($n$) and $\varepsilon$ can not be treated on equal footing. Since these two spacetimes (JP and KTN) are fundamentally different in nature, it is very much expected that the influences of these spacetime in describing the accreting dynamics will not be identical. Hence, the study of the accretion flow properties around them seems important and essential, particularly as case by case manner. In this work, we have done exactly the same and accordingly, we realize that $\varepsilon$ in JP metric and $n$ in KTN metric play similar role (yet not identical) in regulating the accretion solutions. Moreover, beyond a limiting value of $\varepsilon$, the nature of flow solution does not comply with the black hole (BH) background, instead it resembles to the central source that is yielded as naked singularity (NS), which perhaps indicates the following possible implications. \citet[]{Bambhaniya_et_al_2021} reported that for $\theta=\pi/2$, the horizon ceases to exist for a given set of $(a_{\rm k} \leq 1, \varepsilon)$, while horizon exists far from the equatorial plane. Hence, the object is naked only along $\theta=\pi/2$ plane and since the present work is carried out on this plane, it is natural to obtain NS solutions. On the contrary, KTN spacetime results NS $(a_{\rm k} > 1, n)$ irrespective of $\theta$ values. This yields a possible conjecture that the accretion dynamics seem to depend on the existence of horizon on the disk equatorial plane and possibly it dose not crucially depends on the existence or non-existence of horizon on $\theta \ne \pi/2$ planes. Needless to mention that the above findings are realized from the present study only, although we infer that this suggestive conjecture needs further investigation for its conclusiveness.

Finally, we mention that the present study does not include any dissipation processes, namely viscosity, radiative cooling etc. Moreover, we neglect magnetic fields as well for simplicity. All these physical processes are indeed relevant in accretion disc, however, implementation of these processes are beyond the scope of the present work. We plan to take up these issues in our future works and will be reported elsewhere.

\section*{Data availability statement}

The data underlying this article will be available with reasonable request.

\section*{Acknowledgement}

Authors thank the anonymous reviewer for constructive comments and useful suggestions that help to improve the quality of the paper. The work of SP is supported by University Grants Commission (UGC), Government of India, under the scheme Junior Research Fellowship (JRF). BRM is supported by a START-UP RESEARCH GRANT (No. SG/PHY/P/BRM/01) from the Indian Institute of Technology Guwahati, India. The work of SD is supported by the Science and Engineering Research Board (SERB) of India through grant MTR/2020/000331.

\appendix
\begin{widetext}
\section{Derivation of $\lowercase{\boldmath{\frac{dv}{dr}}}$ at the critical point $\lowercase{\boldmath{r_{c}}}$}
\label{appe:appendix}
After applying the l$'$H\^{o}pital's rule, we get the radial velocity gradient at the critical point as
\begin{equation*}
\label{eq:dv/dr_r_c1}
\frac{dv}{dr}\bigg\vert_{c} = \frac{-B\pm\sqrt{B^{2}-4AC}}{2A},
\end{equation*}
where the explicit form of the quantities $A$, $B$ and $C$ are calculated as follows:
\begin{equation*}
A = \gamma_{v}^{2}\left[1 + \frac{2C_{s}^{2}}{(\Gamma + 1)v^{2}}\right] + \frac{4\gamma_{v}^{4}\Theta}{v^{2}(2N + 1)}\frac{\partial}{\partial \Theta}\left( \frac{C_{s}^{2}}{\Gamma + 1}\right),
\end{equation*}
\begin{equation*}
\begin{split}
B  = \frac{8\gamma_{v}^{2}\Theta}{v(2N + 1)}\left[N_{11} + N_{12} - \frac{3\varepsilon}{2r^{4}}N_{13} \right]  \frac{\partial}{\partial \Theta}\left( \frac{C_{s}^{2}}{\Gamma + 1}\right),
\end{split}
\end{equation*}
\begin{equation*}
\begin{split}
C = \frac{d^{2}\Phi^\text{eff}}{dr^{2}} & + \frac{4\Theta}{2N + 1}\left[N_{11} + N_{12} - \frac{3\varepsilon}{2r^{4}}N_{13} \right]^{2} \frac{\partial}{\partial \Theta}\left( \frac{C_{s}^{2}}{\Gamma + 1}\right) \\& - \frac{2C_{s}^{2}}{\Gamma + 1}\left[N_{11}^{\prime} + N_{12}^{\prime} + N_{14} + \frac{3a_{k}^{2}\varepsilon}{2r^{4}(\Delta + a_{k}^{2}h)^{2}}\left(\Delta ^{'} - \frac{3a_{k}^{2}\varepsilon}{r^{4}}\right) \right],
\end{split}
\end{equation*}
\begin{equation*}
N_{11} = \frac{5}{2r} + \frac{r-a_{k}^{2}(1+h)}{r(\Delta + a_{k}^{2}h)}, N_{12} = -\frac{1}{2F}\frac{dF}{dr}, N_{13} = \frac{1}{1+h} + \frac{a_{k}^{2}}{\Delta + a_{k}^{2}h},
\end{equation*}
\begin{equation*}
N_{14} = \frac{6\varepsilon}{r^{5}(1+h)} - \frac{9\varepsilon ^{2}}{2r^{8}(1+h)^{2}} + \frac{6a_{k}^{2}\varepsilon}{r^{5}(\Delta + a_{k}^{2}h)},
\end{equation*}
\begin{equation*}
\begin{split}
\frac{\partial}{\partial\Theta}\left(\frac{C_{s}^{2}}{\Gamma + 1}\right) = \frac{2\Gamma}{(f+2\Theta)(\Gamma + 1)} - \frac{2\Theta\Gamma(2 + f^{\prime})}{\left(f+2\Theta)^{2}(\Gamma + 1)\right)} - \frac{2\Theta\Gamma\Gamma^{\prime}}{(f + 2\Theta)(\Gamma + 1)^{2}} + \frac{2\Theta\Gamma^{\prime}}{(f+2\Theta)(\Gamma + 1)},
\end{split}
\end{equation*}
\begin{equation*}
f^{\prime} = \frac{\partial f}{\partial\Theta}=  N_{1} + N_{2}, \hspace{0.2cm} \Gamma^{\prime} = \frac{\partial\Gamma}{\partial\Theta} = - 2 \frac{N_{1}^{\prime} + N_{2}^{\prime}}{(N_{1} + N_{2})^{2}},
\end{equation*}
\begin{equation*}
N_{1} = \left[\frac{9 \Theta }{3 \Theta +2}-\frac{3\Theta(9 \Theta +3)}{(3 \Theta +2)^2}+\frac{9 \Theta +3}{3 \Theta +2}\right] (2-\xi ), N_{2} = \xi  \left[\frac{9 \Theta }{3 \Theta +\frac{2}{\chi }}-\frac{3 \Theta  \left(9 \Theta +\frac{3}{\chi }\right)}{\left(3 \Theta +\frac{2}{\chi }\right)^2}+\frac{9 \Theta +\frac{3}{\chi }}{3 \Theta +\frac{2}{\chi }}\right],
\end{equation*}
\begin{equation*}
\begin{split}
N_{1}^{\prime} = \frac{\partial N_{1}}{\partial\Theta} = (2-\xi)\left[\frac{18\Theta(9 \Theta +3)}{(3 \Theta +2)^3}-\frac{54 \Theta }{(3 \Theta +2)^2}+\frac{18}{3 \Theta +2}-\frac{6 (9 \Theta +3)}{(3 \Theta +2)^2}\right],
\end{split}
\end{equation*}
\begin{equation*}
\begin{split}
N_{2}^{\prime}=\frac{\partial N_2}{\partial\Theta} = \xi \left[\frac{18 \Theta  \left(9 \Theta +\frac{3}{\chi }\right)}{\left(3 \Theta +\frac{2}{\chi }\right)^3}-\frac{54 \Theta }{\left(3 \Theta +\frac{2}{\chi }\right)^2}+\frac{18}{3 \Theta +\frac{2}{\chi }}-\frac{6 \left(9 \Theta +\frac{3}{\chi }\right)}{\left(3 \Theta +\frac{2}{\chi }\right)^2}\right],
\end{split}
\end{equation*}
\begin{equation*}
\frac{d^{2}\Phi_\text{eff}}{dr^{2}} = \frac{1}{2}\left( N_{21} + N_{22} + N_{23} -\frac{1}{2r^{2}} \right), N_{21} = \left( \frac{N_{111}^{\prime}}{N_{111}} \right)^{2}-\frac{N_{111}^{\prime\prime}}{N_{111}},
\end{equation*}
\begin{equation*}
N_{22} = \frac{h^{\prime\prime}}{1+h} - \left(\frac{h^{\prime}}{1+h}\right)^{2}, N_{23} = \frac{2+a_{k}^{2}h^{\prime\prime}}{\Delta + a_{k}^{2}h} - \left(\frac{\Delta^{\prime}+a_{k}^{2}h}{\Delta + a_{k}^{2}h}\right)^{2}, \Delta^{\prime} =\frac{d\Delta}{dr} = 2(r-1), h^{\prime} = \frac{dh}{dr} = -\frac{3\varepsilon}{r^{4}},
\end{equation*}
\begin{equation*}
\begin{split}
h^{\prime\prime}= \frac{dh^{\prime}}{dr}= \frac{12\varepsilon}{r^{5}}, N_{111} = r^{3}+a_{k}^{2}(r+2)(1+h)-4a_{k}\lambda(1+h)-\lambda_{k}^{2}(r-2)(1+h),
\end{split}
\end{equation*}
\begin{equation*}
\begin{split}
N_{111}^{\prime} = \frac{dN_{111}}{dr} = 3r^{2} + a_{k}^{2}(r + 2)h^{\prime} + a_{k}^{2}(1 + h) - 4a_{k}\lambda h^{\prime} - \lambda^{2}(1 + h) - \lambda^{2}(r-2)h^{\prime},
\end{split}
\end{equation*}
\begin{equation*}
\begin{split}
N_{111}^{\prime\prime} = \frac{dN_{111}^{\prime}}{dr} = 6r + \left[ a_{k}^{2}(r+2) -4a_{k}\lambda - \lambda^{2}(r-2) \right]h^{\prime\prime} + \left( 2a_{k}^{2}-\lambda^{2}\right)h^{\prime}.
\end{split}
\end{equation*}
Here, all the quantities have their usual meaning.
\end{widetext}

\bibliographystyle{apsrev}
\bibliography{references} 

\end{document}